\documentclass{article}

\usepackage{amssymb,latexsym,amsmath}

\usepackage[pdftex]{graphicx}

\usepackage{hyperref}

\begin{document}

\pagestyle{plain}

\newtheorem{theorem}{Theorem}[section]

\newtheorem{proposition}[theorem]{Proposition}

\newtheorem{lema}[theorem]{Lemma}

\newtheorem{corollary}[theorem]{Corollary}

\newtheorem{definition}[theorem]{Definition}

\newtheorem{remark}[theorem]{Remark}

\newtheorem{exempl}{Example}[section]

\newenvironment{exemplu}{\begin{exempl}  \em}{\hfill $\square$

\end{exempl}}

\newcommand{\ea}{\mbox{{\bf a}}}

\newcommand{\eu}{\mbox{{\bf u}}}

\newcommand{\ueu}{\underline{\eu}}

\newcommand{\ueo}{\overline{u}}

\newcommand{\oeu}{\overline{\eu}}

\newcommand{\ew}{\mbox{{\bf w}}}

\newcommand{\ef}{\mbox{{\bf f}}}

\newcommand{\eF}{\mbox{{\bf F}}}

\newcommand{\eC}{\mbox{{\bf C}}}

\newcommand{\en}{\mbox{{\bf n}}}

\newcommand{\eT}{\mbox{{\bf T}}}

\newcommand{\eL}{\mbox{{\bf L}}}

\newcommand{\eR}{\mbox{{\bf R}}}

\newcommand{\eV}{\mbox{{\bf V}}}

\newcommand{\eU}{\mbox{{\bf U}}}

\newcommand{\ev}{\mbox{{\bf v}}}

\newcommand{\eve}{\mbox{{\bf e}}}

\newcommand{\uev}{\underline{\ev}}

\newcommand{\eY}{\mbox{{\bf Y}}}

\newcommand{\eK}{\mbox{{\bf K}}}

\newcommand{\eP}{\mbox{{\bf P}}}

\newcommand{\eS}{\mbox{{\bf S}}}

\newcommand{\eJ}{\mbox{{\bf J}}}

\newcommand{\eB}{\mbox{{\bf B}}}

\newcommand{\eH}{\mbox{{\bf H}}}

\newcommand{\leb}{\mathcal{ L}^{n}}

\newcommand{\eI}{\mathcal{ I}}

\newcommand{\eE}{\mathcal{ E}}

\newcommand{\hen}{\mathcal{H}^{n-1}}

\newcommand{\eBV}{\mbox{{\bf BV}}}

\newcommand{\eA}{\mbox{{\bf A}}}

\newcommand{\eSBV}{\mbox{{\bf SBV}}}

\newcommand{\eBD}{\mbox{{\bf BD}}}

\newcommand{\eSBD}{\mbox{{\bf SBD}}}

\newcommand{\ecs}{\mbox{{\bf X}}}

\newcommand{\eg}{\mbox{{\bf g}}}

\newcommand{\paromega}{\partial \Omega}

\newcommand{\gau}{\Gamma_{u}}

\newcommand{\gaf}{\Gamma_{f}}

\newcommand{\sig}{{\bf \sigma}}

\newcommand{\gac}{\Gamma_{\mbox{{\bf c}}}}

\newcommand{\deu}{\dot{\eu}}

\newcommand{\dueu}{\underline{\deu}}

\newcommand{\dev}{\dot{\ev}}

\newcommand{\duev}{\underline{\dev}}

\newcommand{\weak}{\stackrel{w}{\approx}}

\newcommand{\mild}{\stackrel{m}{\approx}}

\newcommand{\lrightarrow}{\stackrel{L}{\rightarrow}}

\newcommand{\rrightarrow}{\stackrel{R}{\rightarrow}}

\newcommand{\strong}{\stackrel{s}{\approx}}

\newcommand{\weakdown}{\rightharpoondown}

\newcommand{\opg}{\stackrel{\mathfrak{g}}{\cdot}}

\newcommand{\opunu}{\stackrel{1}{\cdot}}
\newcommand{\opdoi}{\stackrel{2}{\cdot}}

\newcommand{\opn}{\stackrel{\mathfrak{n}}{\cdot}}
\newcommand{\opx}{\stackrel{x}{\cdot}}

\newcommand{\tr}{\ \mbox{tr}}

\newcommand{\Ad}{\ \mbox{Ad}}

\newcommand{\ad}{\ \mbox{ad}}

\renewcommand{\contentsname}{ }

\title{Local and global moves on locally  planar  trivalent graphs, lambda calculus and $\lambda$-Scale}

\author{Marius Buliga \\ 
\\
Institute of Mathematics, Romanian Academy \\
P.O. BOX 1-764, RO 014700\\
Bucure\c sti, Romania\\
{\footnotesize Marius.Buliga@imar.ro}}

\date{This version: 02.07.2012}

\maketitle

\begin{abstract}
We give a description of local and global moves on a class of locally  planar trivalent graphs and we show that it contains  $\lambda$-Scale calculus, therefore in particular untyped  lambda calculus.  Surprisingly, the beta reduction rule comes from a local "sewing" transformation of trivalent locally  planar graphs. 
\end{abstract}

\section{Introduction}

$\lambda$-Scale calculus, introduced in \cite{lambdascale}, is a formalism which contains both untyped lambda calculus and emergent algebras. It has been conceived in order to give a rigorous meaning to "computing with space" \cite{buligachora}, i.e. to give a precise meaning to  computation associated to emergent algebras. 

Emergent algebras \cite{buligairq} \cite{buligabraided} are a distillation of differential calculus in metric spaces with dilations \cite{buligadil1}. This class of metric spaces contain the "classical" riemannian manifolds, as well as fractal like spaces as Carnot groups or, more general, sub-riemannian or Carnot-Carath\'eodory spaces, but also all kind of spaces obtained from differential geometric constructions under the constraint of low regularity. 

In the paper \cite{buligachora} we proposed a formalism of decorated tangle diagrams for emergent algebras and we called "computing with space" the various manipulations of these diagrams with geometric content. Nevertheless, in that paper we were not able to give a precise sense of the use of the word "computing". We speculated, by using analogies from studies of the visual system, especially the "Brain a geometry engine" paradigm of Koenderink \cite{koen}, that, in order for the visual front end of the brain to reconstruct the visual space in the brain, there should be a kind of "geometrical computation" in the neural network of the brain 
akin to the manipulation of decorated tangle diagrams described in our paper. 

Tangle diagrams decorated by quandles or racks  are a well known tool in knot theory \cite{fennrourke} \cite{joyce}. Emergent algebras are a generalization of quandles, namely an emergent algebra is a family of idempotent right quasigroups indexed by the elements of an abelian group, while quandles are self-distributive idempotent right quasigroups. What was missing in this picture was only a notion of computation. It is notable to mention the work of Kauffman \cite{kauf}, where the author uses knot diagrams for representing combinatory logic, thus untyped lambda calculus. Moreover, there are ways of transforming knot and tangle diagrams into trivalent graphs, as explained in \cite{buligachora} section 3.1. 

In this paper we give a description of $\lambda$-Scale calculus, therefore in particular of untyped  lambda calculus, in terms of transformations of a  class of locally  planar trivalent graphs. It is very intriguing that the $\beta$-reduction rule of lambda calculus  appear as a local transformation of locally  planar trivalent graphs.

\paragraph{Acknowledgement.} This work was supported by a grant of the Romanian National Authority for Scientific Research, CNCS – UEFISCDI, project number 
PN-II-ID-PCE-2011-3-0383.

\section{Graphs and moves}
\label{gra}

An oriented graph is a pair $(V,E)$, with $V$ the set of nodes and $E \subset V \times V$ the set of edges.  Let us denote by $\displaystyle \alpha: V  \rightarrow 2^{E}$ the map which associates to any node $N \in V$ the set of adjacent edges $\alpha(N)$. In this paper we work with locally planar graphs with  decorated nodes, i.e. we shall attach to a graph $(V,E)$ supplementary information: 
\begin{enumerate}
\item[-] a function $f: V \rightarrow A$ which associates to any node $N \in V$ an element 
of the "graphical alphabet" $A$ (see definition \ref{defalp}), 
\item[-] a cyclic order of $\alpha(N)$ for any $N \in V$, which is equivalent to giving a  local embedding of the node $N$ and edges adjacent to it into the plane. 
\end{enumerate}

We shall construct a set of locally  planar graphs with decorated nodes, starting from a graphical alphabet of  elementary graphs. On the set of graphs we shall define  local transformations, or moves.  Global moves or conditions will be then introduced. 

\begin{definition}
The graphical alphabet contains the elementary graphs, or gates, denoted by $\lambda$, $\Upsilon$, $\curlywedge$, $\top$, and for any element $\varepsilon$ of the commutative group $\Gamma$, a graph denoted by  $\bar{\varepsilon}$. Here are the elements of the graphical alphabet: 
\begin{enumerate}
\item[] $\lambda$ graph \hspace{1.cm} \includegraphics[width=20mm]{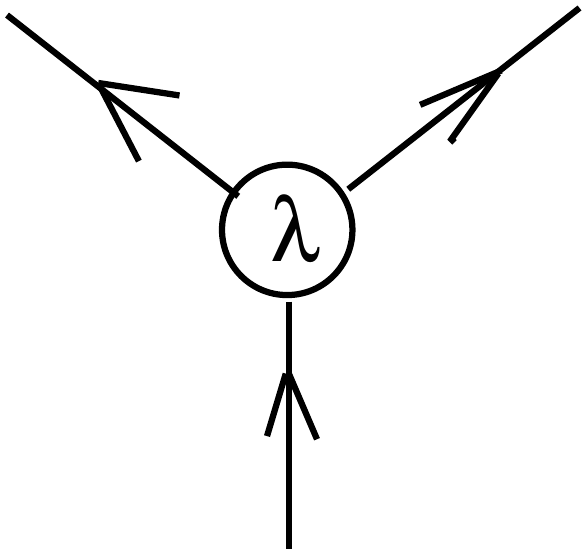}, \hspace{2.cm}
$\Upsilon$ graph \hspace{1.cm}\includegraphics[width=20mm]{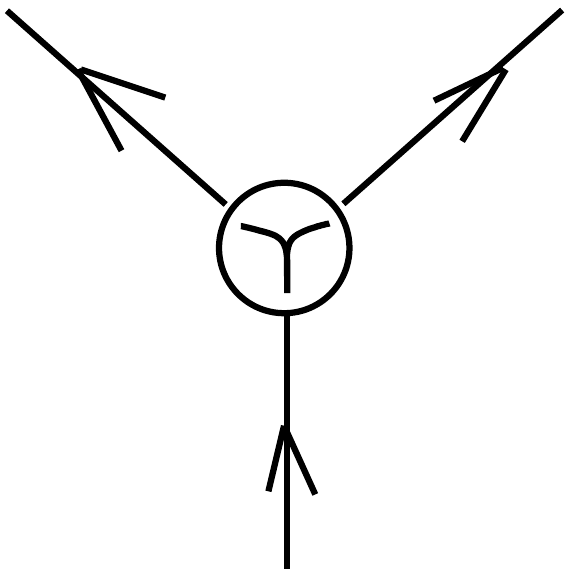}, 

\item[] $\curlywedge$ graph \hspace{1.cm}\includegraphics[width=15mm]{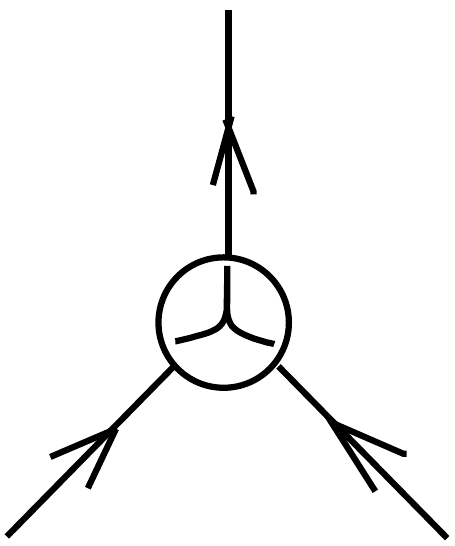}\hspace{1.cm}, \hspace{2.cm} $\bar{\varepsilon}$ graph \hspace{1.cm}\includegraphics[width=15mm]{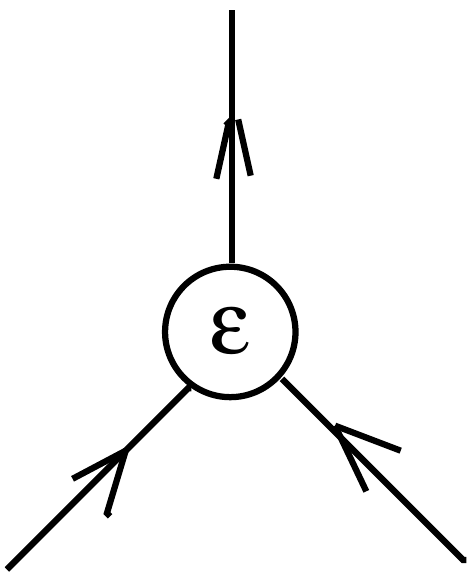}\hspace{.2cm}, 
\item[] $\top$ graph \hspace{1.cm}\includegraphics[width=8mm]{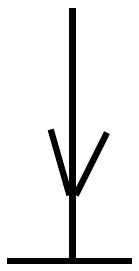}\hspace{1.cm}.
\end{enumerate}
With the exception of the $\top$, all other elementary graphs have three edges. The graph  $\top$ has only one edge. 
\label{defalp}
\end{definition}

\paragraph{1. The set GRAPH.} We construct the set of graphs $GRAPH$ over the graphical alphabet by grafting edges of a finite number of copies of the elements  of the graphical alphabet. 

\begin{definition} 
$GRAPH$ is the set of graphs obtained by grafting edges of a finite number of copies of the elements  of the graphical alphabet. During the grafting procedure, we start from a set of gates and we add, one by one, a finite number of gates, such that, at any step,  any edge of any element of any gate is grafted on any other free  edge (i.e. not already grafted to other edge) of the graph, with the condition  that they have  the same orientation.

For any node of the graph, the local embedding into the plane is given by the element of the graphical alphabet which decorates it. 

The set of free edges of a graph $G \in GRAPH$ is named the set of leaves $L(G)$. Technically, one may imagine that we complete the graph $G \in GRAPH$ by adding to the free extremity of any free edge a decorated node, called "leaf",  with decoration "IN" or "OUT", depending on the orientation of the respective free edge. The set of leaves $L(G)$ thus decomposes into a disjoint union $L(G) = IN(G) \cup OUT(G)$ of in or out leaves. 
\end{definition}


\paragraph{2. The local moves.} These are transformations of graphs in $GRAPH$ which are local, 
in the sense that any of the  moves apply to a limited part of a graph, keeping the rest of the graph unchanged. 

We may define a local move as a rule of transformation of a graph into another of the following form.

First, a subgraph of a graph $G$ in $GRAPH$ is any collection of nodes and/or edges of $G$. It is not supposed that the mentioned subgraph must  be in $GRAPH$. Also, a collection of some edges of $G$, without any node, count as a subgraph of $G$. Thus, a subgraph of $G$ might be imagined as a subset of the reunion of nodes and edges of $G$. 

For any   natural  number $N$ and any graph $G$ in $GRAPH$, let  
$\displaystyle \mathcal{P}(G,N)$ be the collection  of subgraphs $P$ of the graph $G$ which have the sum of the number of  edges and nodes less than or equal to $N$.

\begin{definition}
A local move has the following form: there is a number $N$ and a condition $C$ which is formulated in terms of graphs which have the sum of the number of  edges and nodes less than or equal to $N$,  such that for any graph $G$ in $GRAPH$ and for any $P \in \mathcal{P}(G,N)$, if $C$ is true for $P$ then transform $P$ into $P'$, where $P'$ is also a graph which have the sum of the number of  edges and nodes less than or equal to $N$. 
\end{definition}

Graphically we shall group the elements of the subgraph,  subjected to the application of the local rule,  into a region  encircled with a dashed closed, simple curve.  The edges which cross the curve (thus connecting the subgraph $P$ with the rest of the graph) will be numbered clockwise. The transformation will affect only the part of the graph which is inside the dashed curve (inside meaning the bounded connected part of the plane which is bounded by the dashed curve) and, after the transformation is performed, the edges of the transformed graph will connect to the graph outside the dashed curve by respecting the numbering of the edges which cross the dashed line.

\paragraph{2.1. Graphic $\beta$ move.} This is the most important move, inspired by the $\beta$-reduction from lambda calculus, see theorem \ref{lambdathm}, see also the ($\beta$*) rule in $\lambda$-Scale calculus  \cite{lambdascale} and theorem \ref{lambdascalethm} here.

\centerline{\includegraphics[width=80mm]{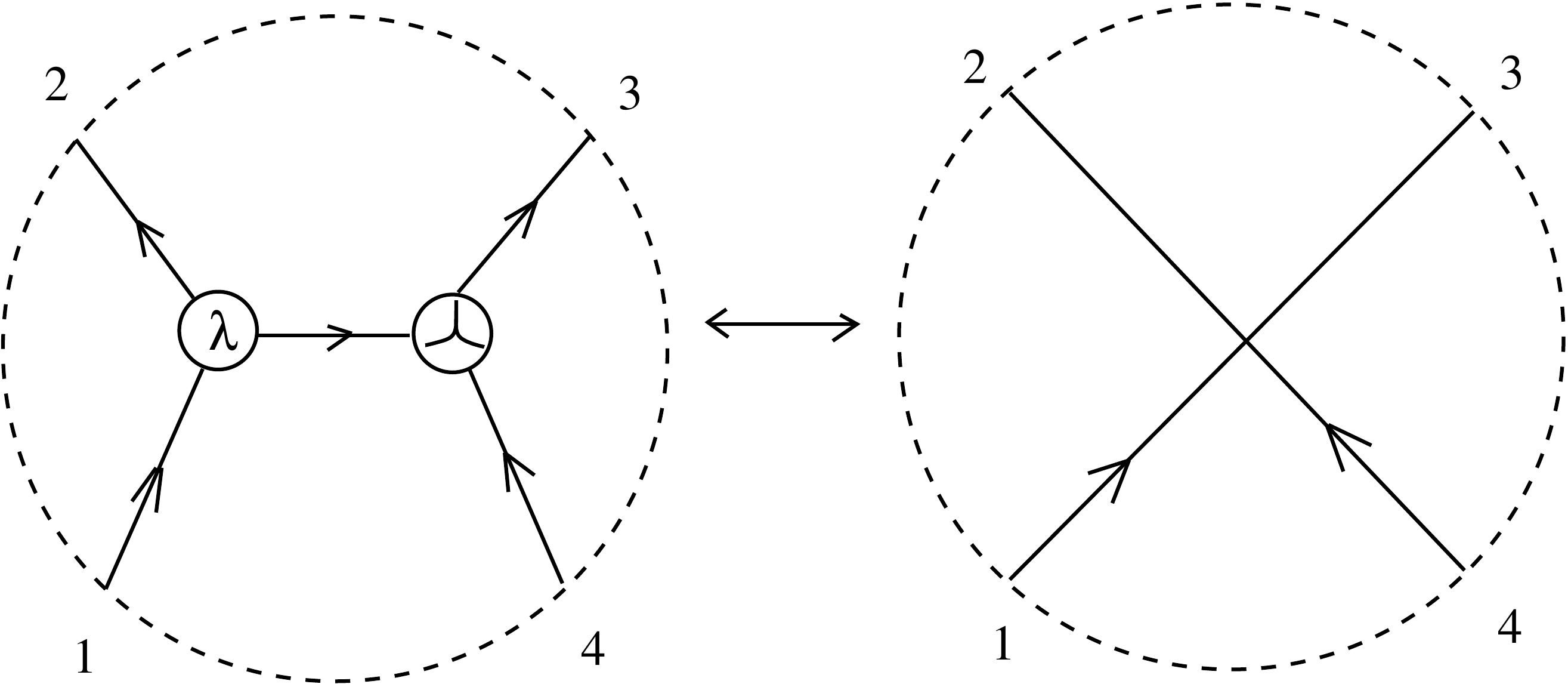}}

This "sewing braids" move will be used also in contexts outside of lambda calculus! It is the most powerful move in this graphic calculus. A primitive form of this move appears as the re-wiring move (W1) (section 3.3, p. 20  and the last paragraph and figure from section 3.4, p. 21 in \cite{buligachora}). 

An alternative notation for this move is the following:

\centerline{\includegraphics[width=80mm]{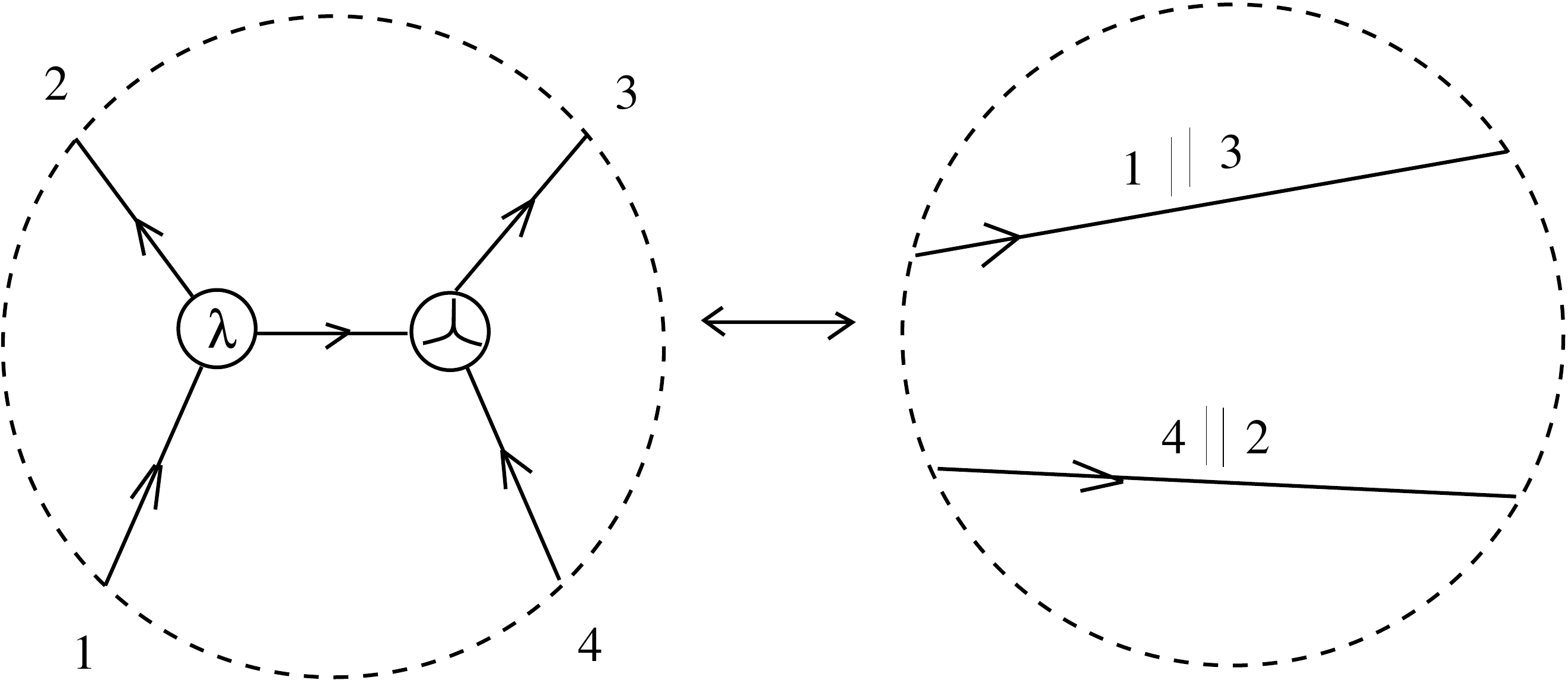}}

\paragraph{2.2. (CO-ASSOC) move.} This is the "co-associativity" move involving the $\Upsilon$ graphs. We think about the  $\Upsilon$ graph as corresponding to a FAN-OUT gate (however, the (CO-ASSOC) alone does not completely characterize $\Upsilon$ as being a FAN-OUT gate). 

\centerline{\includegraphics[width=80mm]{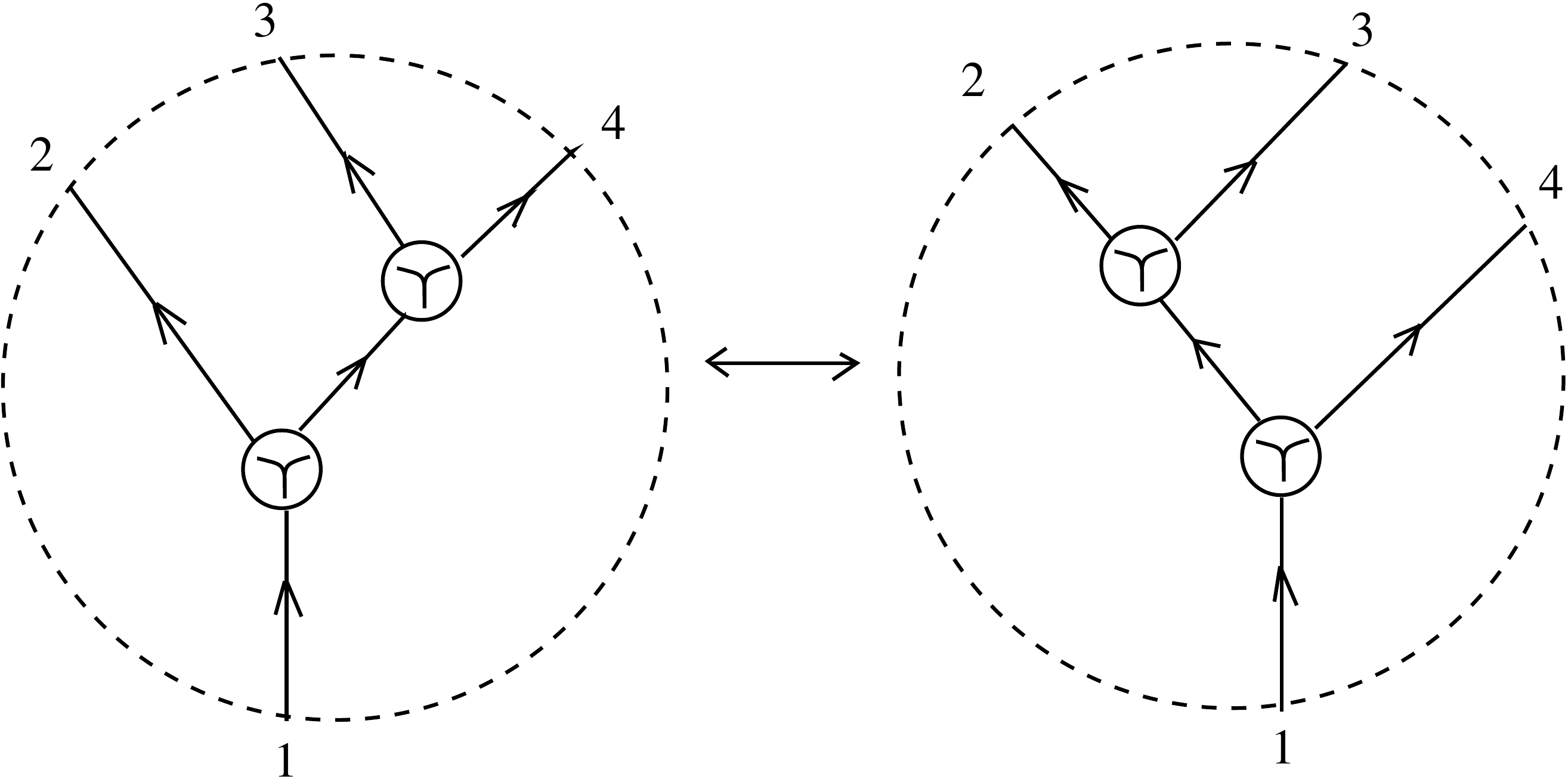}}

\paragraph{2.3. (CO-COMM) move.} This is the "co-commutativity" move involving the $\Upsilon$ gate. It will be not used until the section \ref{secbraid} concerning knot diagrams.

\centerline{\includegraphics[width=80mm]{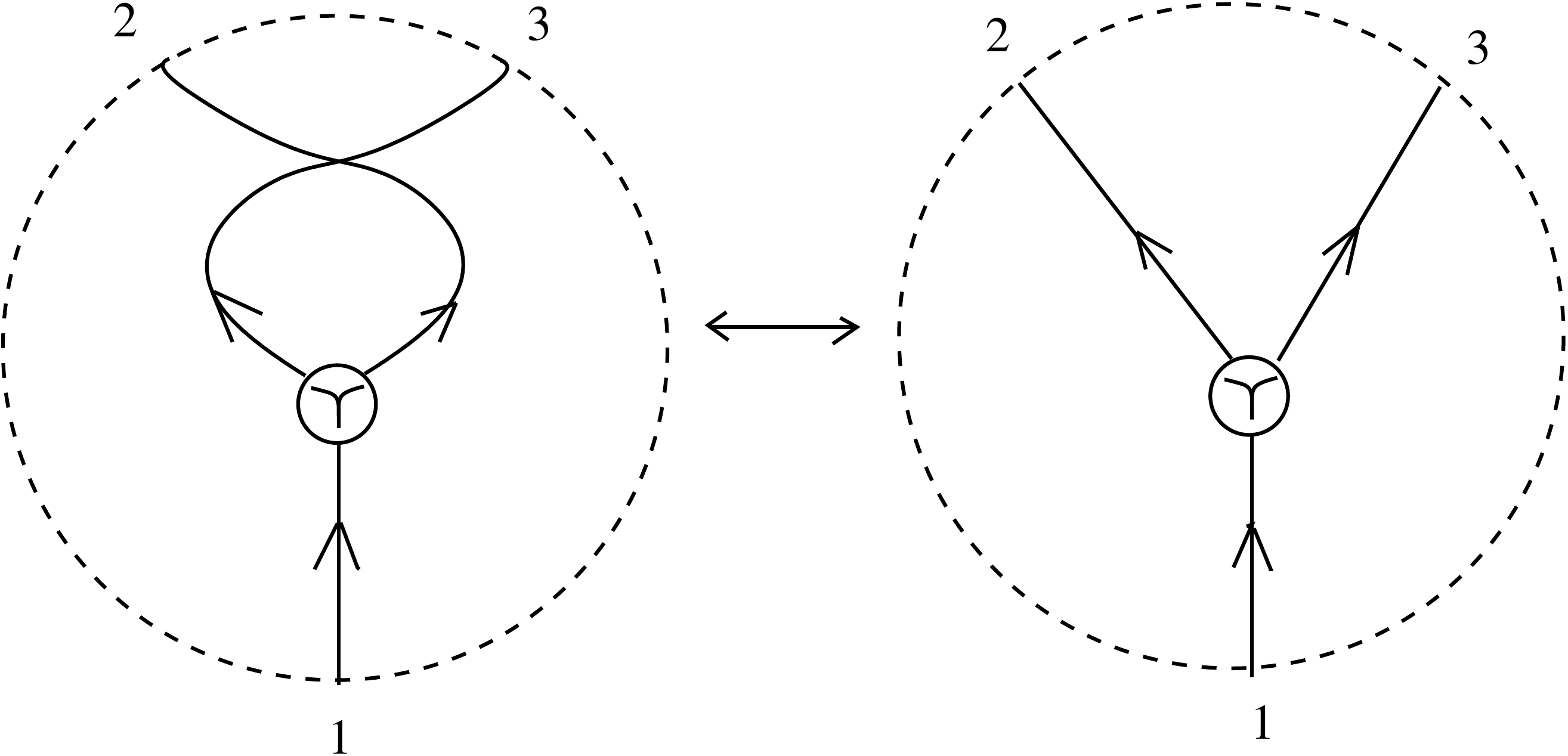}}

\paragraph{2.3. (R1) move.} This corresponds to the Reidemeister I move for emergent algebras. It involves an $\Upsilon$ graph and a $\bar{\varepsilon}$ graph, with $\varepsilon \in \Gamma$. 

\centerline{\includegraphics[width=80mm]{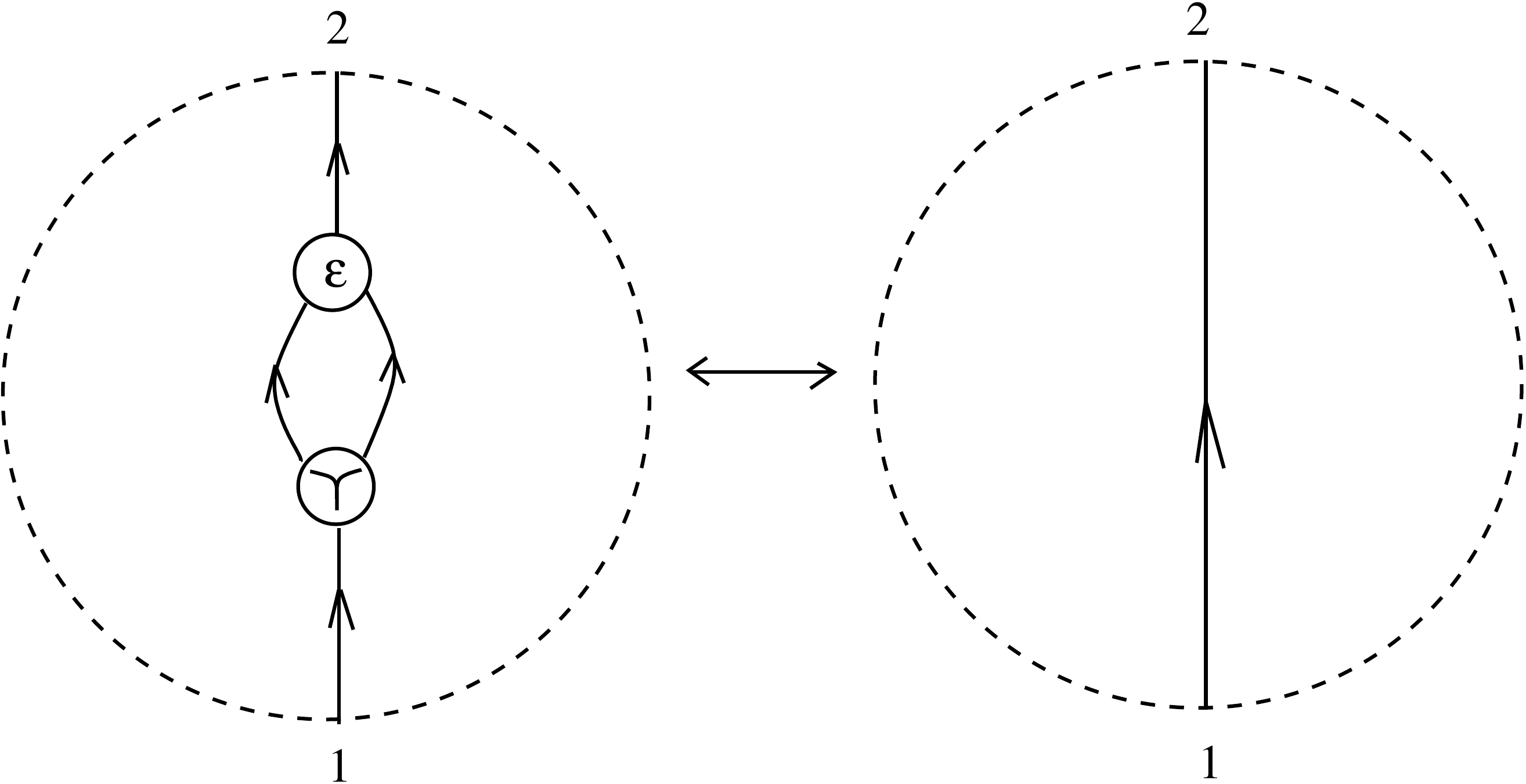}}

It appears also in section 3.4, p. 21 \cite{buligachora}. 

\paragraph{2.4. (R2) move.} This corresponds to the Reidemeister II move for emergent algebras. It involves an $\Upsilon$ graph and two other: a $\bar{\varepsilon}$ and a $\bar{\mu}$ graph, with $\varepsilon, \mu \in \Gamma$.

\centerline{\includegraphics[width=80mm]{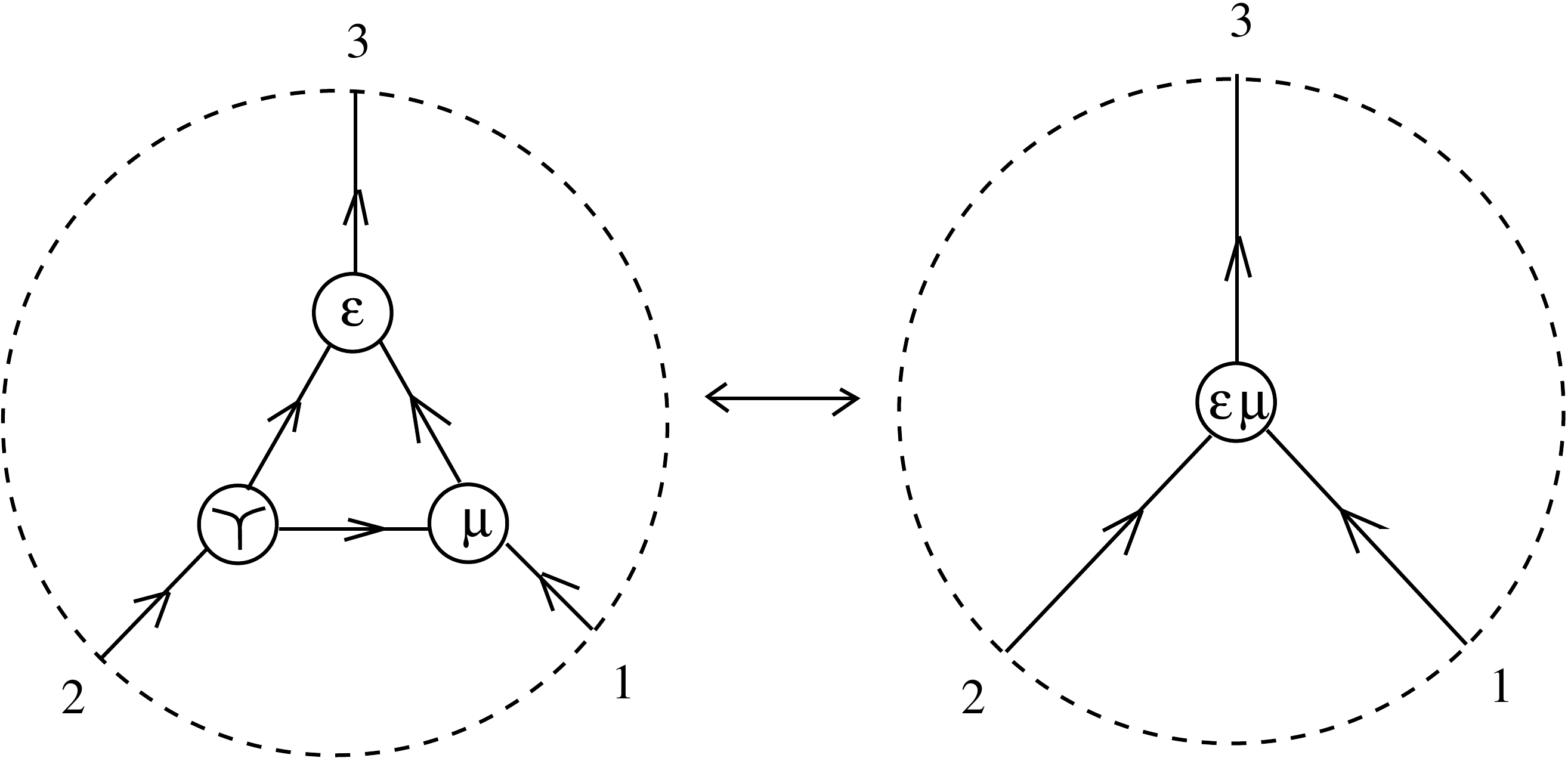}}

This move appears in section 3.4, p. 21 \cite{buligachora}, with the supplementary name "triangle move".

\paragraph{2.5. (ext2) move.} This corresponds to the rule (ext2) from $\lambda$-Scale calculus, it expresses the fact that in emergent algebras the operation indexed with the neutral element $1$ of the group $\Gamma$ has the property $\displaystyle x \circ_{1} y = y$. 

\centerline{\includegraphics[width=80mm]{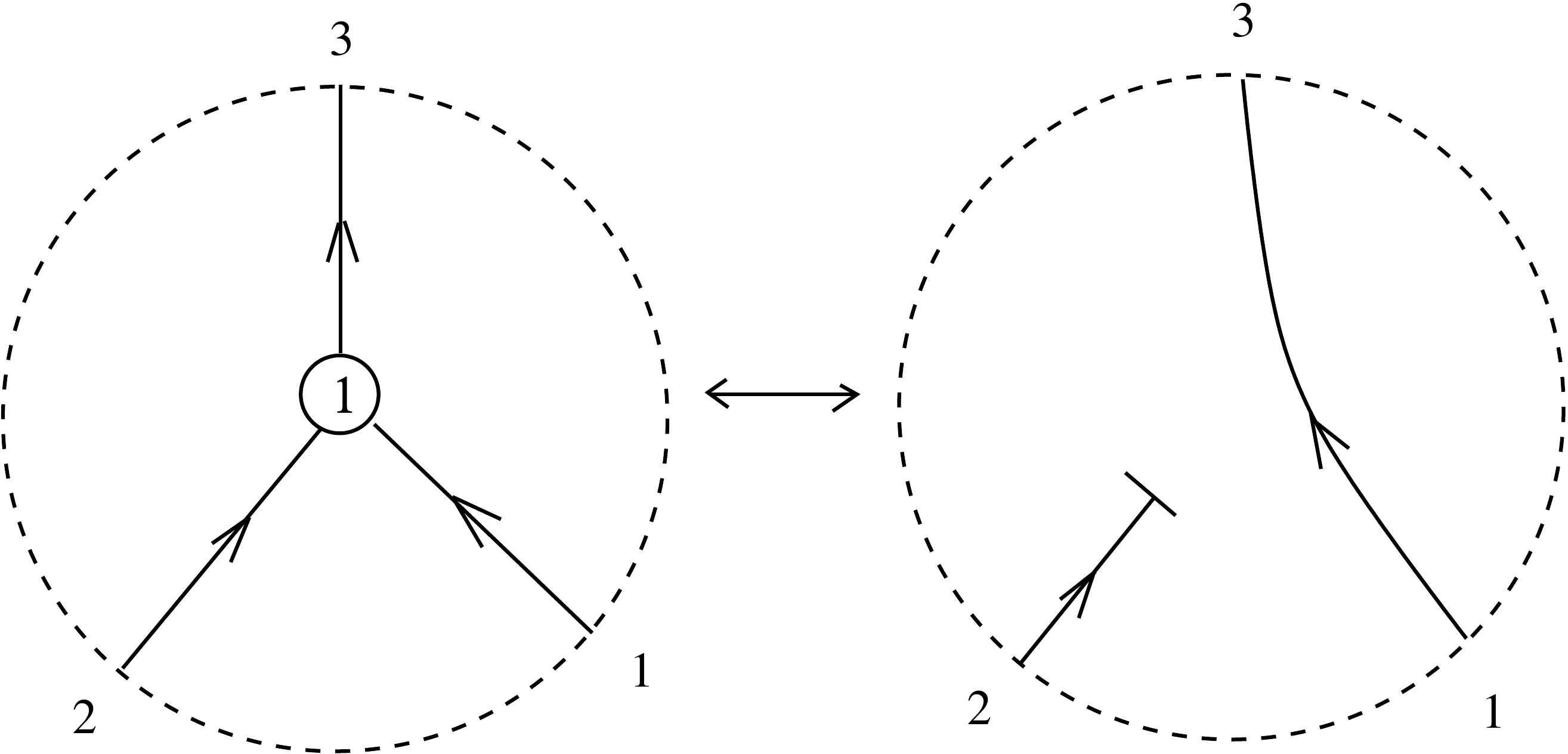}}

\paragraph{2.6. Local pruning.} These are local moves which eliminate "dead" edges. 

\vspace{.5cm}

\centerline{\includegraphics[width=90mm]{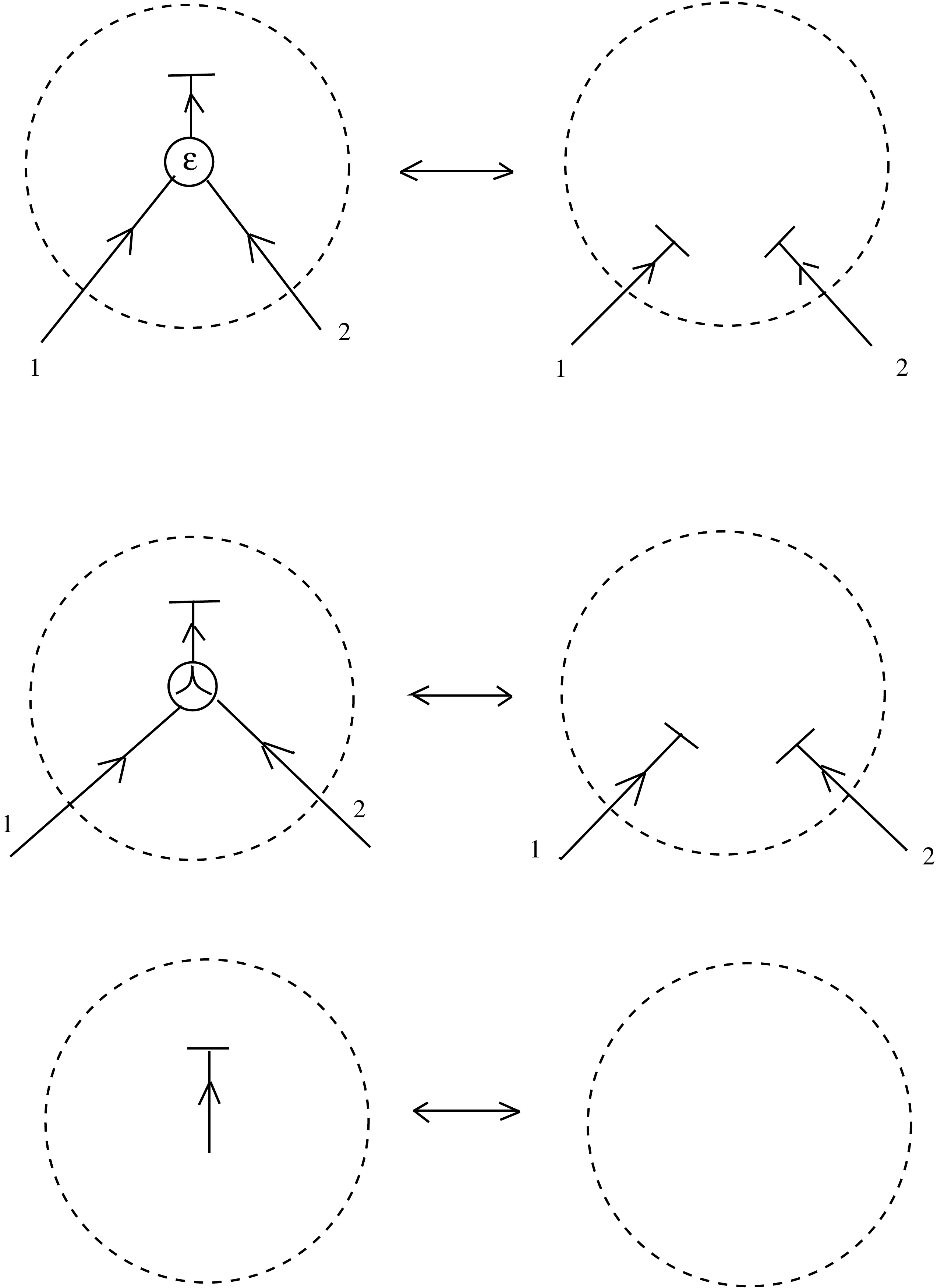}}

\vspace{.5cm}

\centerline{\includegraphics[width=90mm]{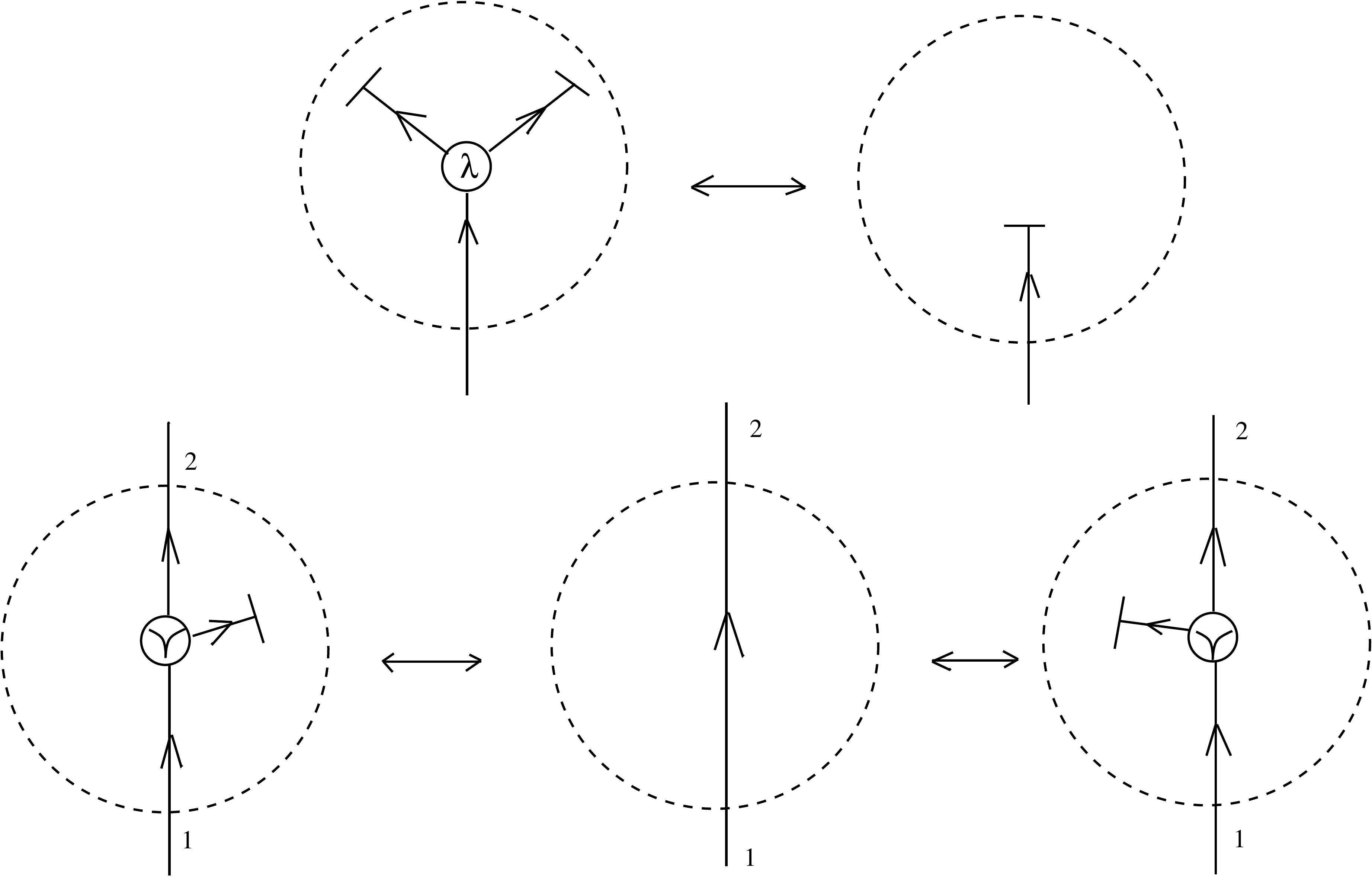}}

\vspace{.5cm}

\paragraph{Global  moves or conditions.} Global  moves are those which are not local, either because the condition $C$ applies to parts of the graph which may have an arbitrary large sum or edges plus nodes, or because after the move the graph $P'$ which replaces the graph $P$ has an arbitrary large sum or edges plus nodes.

\paragraph{2.7. (ext1) move.} This corresponds to the rule (ext1) from $\lambda$-Scale calculus, or to $\eta$-reduction in lambda calculus. It involves a $\lambda$ graph (think about the $\lambda$ abstraction operation in lambda calculus) and a $\curlywedge$ graph (think about   the application operation in lambda calculus). 

The rule is: if there is no oriented path from "2" to "1" outside of the dashed curve, then 
the following move can be performed. 
 
\centerline{\includegraphics[width=80mm]{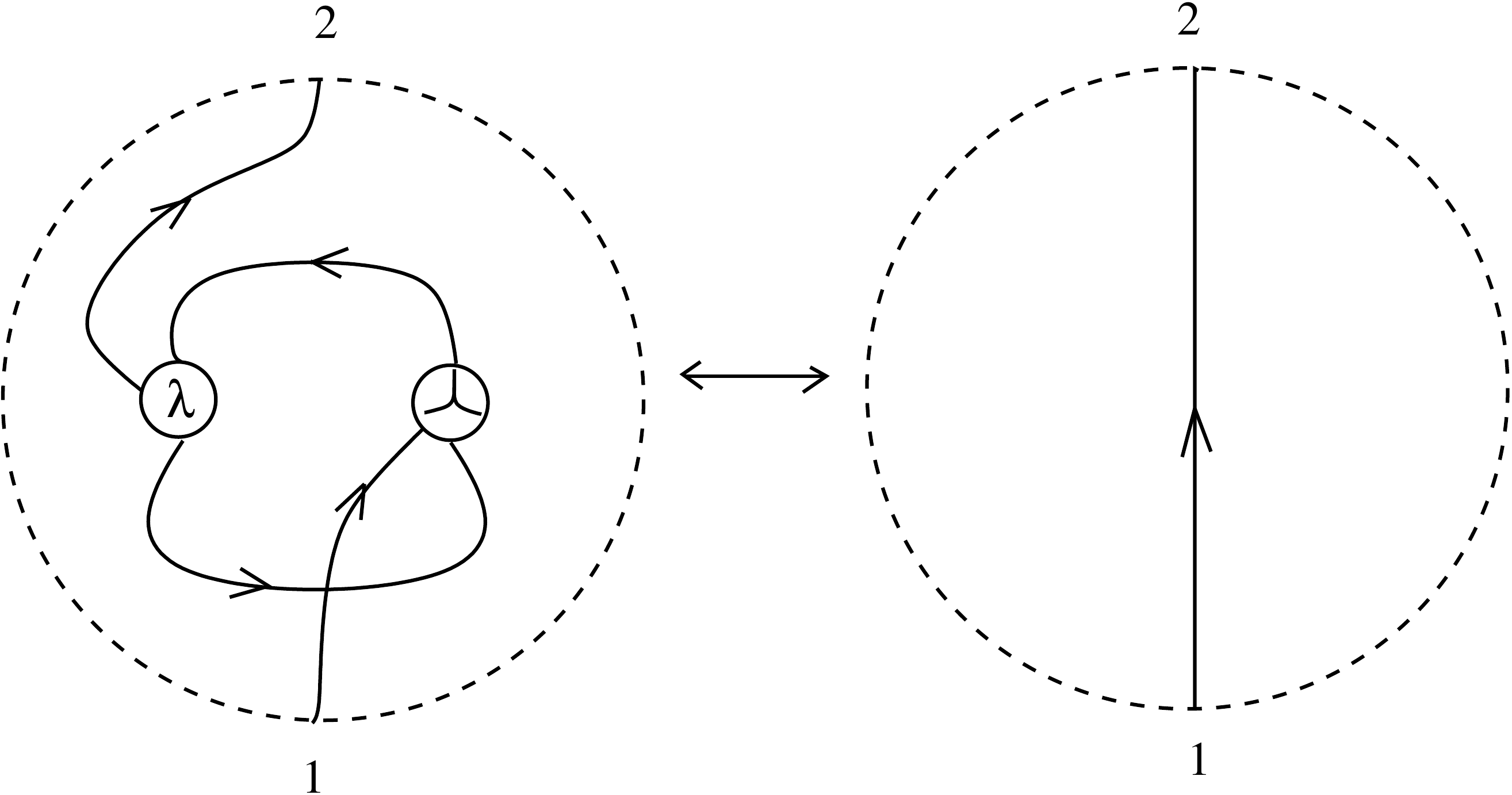}}

\paragraph{2.8. (Global FAN-OUT) move.} This is a global move, because it consists in replacing (under certain circumstances) of a graph by two copies of that graph. 

The rule is: if a graph in $G \in GRAPH$ has a $\Upsilon$ bottleneck, that is if we can find a sub-graph $A \in GRAPH$ connected to the rest of the graph $G$ only through a $\Upsilon$ gate, then we can perform the move  explained in the next figure, from the left to the right.

 Conversely, if in the graph $G$ we can find two identical subgraphs (denoted by $A$), which are in $GRAPH$, which have no edge connecting one with another and which are connected to the rest of $G$ only through one edge, as in the RHS of the figure, then we can perform the move from the right to the left. 

\vspace{.5cm}

\centerline{\includegraphics[width=80mm]{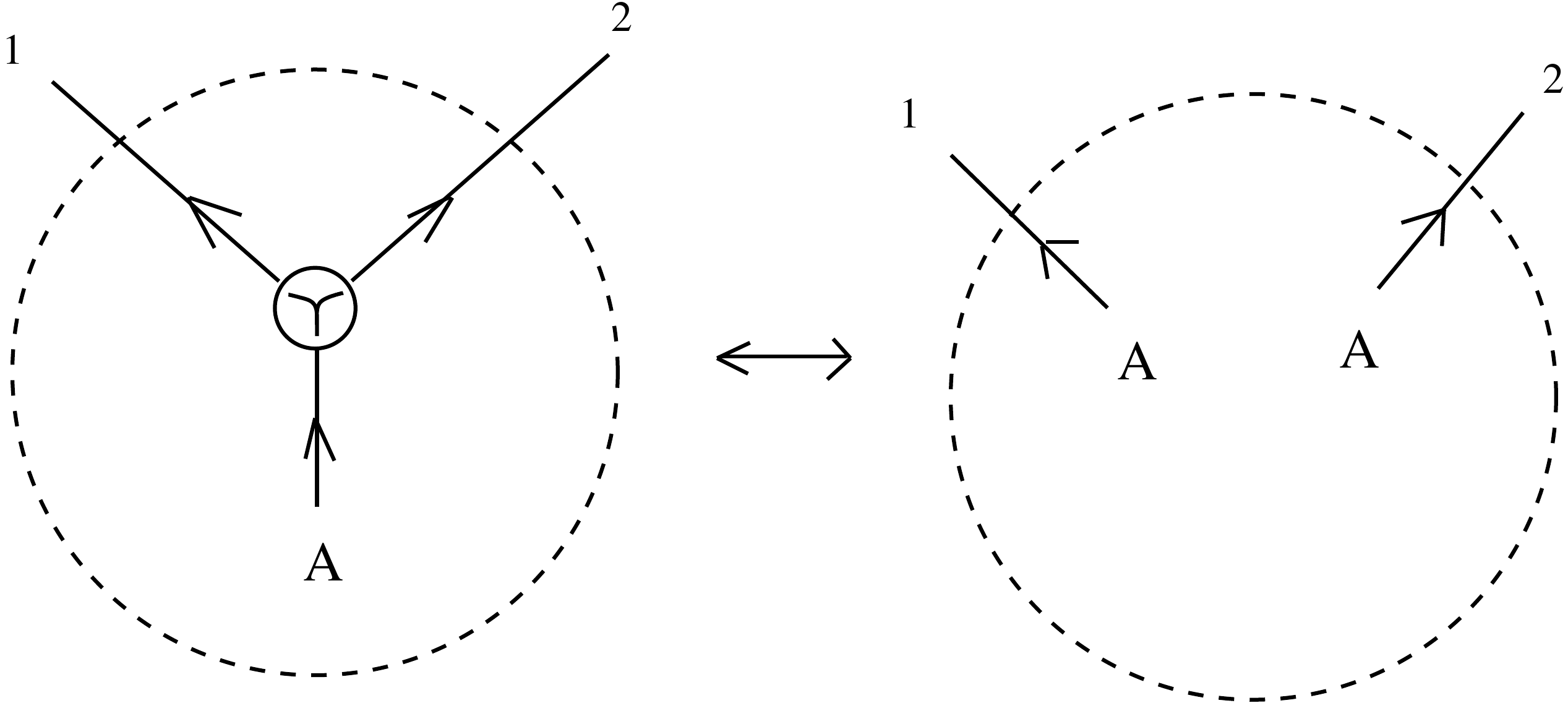}}

\vspace{.5cm}

Remark that (global FAN-OUT) trivially implies (CO-COMM). 

\paragraph{2.9. Global pruning.} This a global move which eliminates "dead" edges.

The rule is: if a graph in $G \in GRAPH$ has a $\top$ ending, that is if we can find a sub-graph $A \in GRAPH$  connected only to a  $\top$ gate, with no edges connecting to the rest of $G$, then we can erase this graph and the respective $\top$ gate.

 Conversely, we may add to a graph $G \in GRAPH$ a graph with a $\top$ ending, not connected with $G$. 

\vspace{.5cm}

\centerline{\includegraphics[width=80mm]{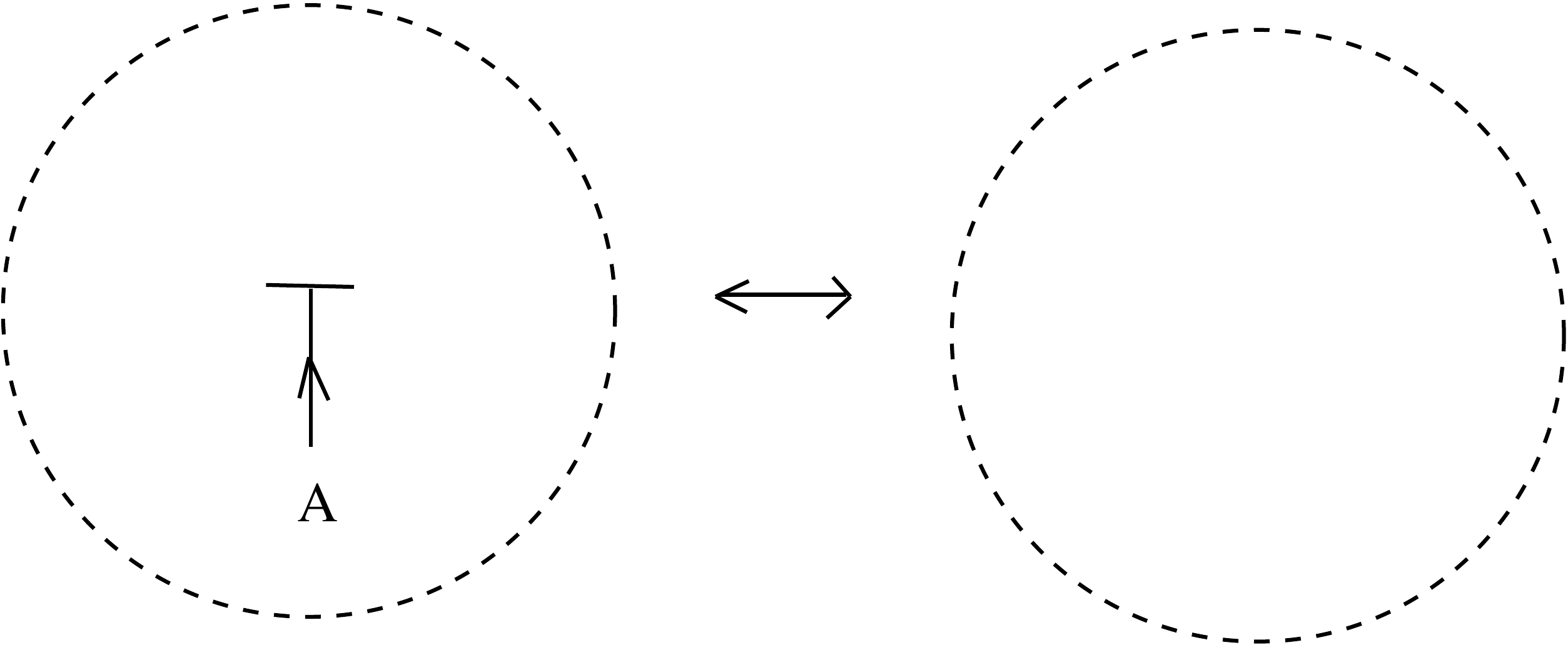}}

\vspace{.5cm}

The global pruning may be needed because of the $\lambda$ gates, which cannot be removed only by local pruning. 

\paragraph{2.10. Elimination of loops.} It is possible that, after using a local or global move, we obtain a graph with an arrow which closes itself, without being connected to any node. 
Here is an example, concerning the application of the graphic $\beta$ move.

\vspace{.5cm}

\centerline{\includegraphics[width=80mm]{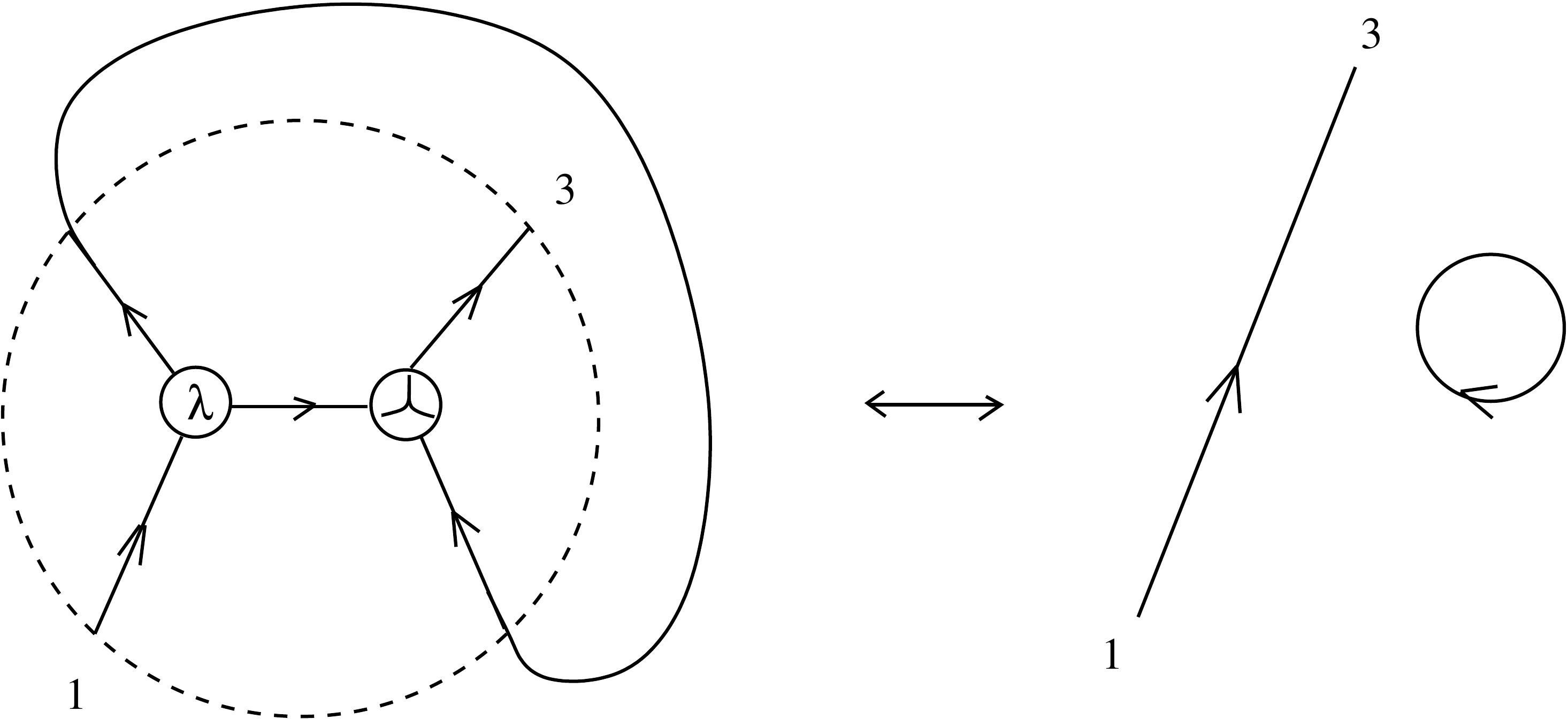}}

\vspace{.5cm}

We shall erase any such loop, by convention. As an illustration, the previous figure, with loop elimination, takes the following form. 

\vspace{.5cm}

\centerline{\includegraphics[width=100mm]{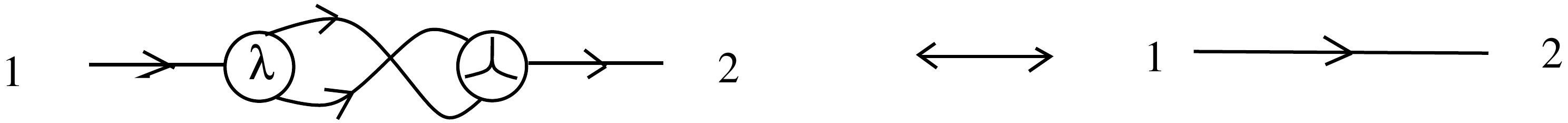}}

\vspace{.5cm}

\paragraph{$\lambda$GRAPHS.} The edges of an elementary graph $\lambda$ can be numbered unambiguously, clockwise, by 1, 2, 3, such that 1 is the number of the entrant edge.

\begin{definition}
A graph $G \in GRAPH$ is a $\lambda$-graph, notation $G \in \lambda GRAPH$, if: 
\begin{enumerate}
\item[-] it does not have $\bar{\varepsilon}$ gates, 
\item[-] for any node $\lambda$ any oriented path in $G$ starting at the edge 2 of this node can be completed to a path which either terminates in a graph $\top$, or else terminates at the edge 1 of this node. 
\end{enumerate}
\end{definition}
The condition $G \in \lambda GRAPH$ is global, in the sense that  in order to decide if $G \in \lambda GRAPH$ we have to examine parts of the graph which may have an arbitrary large sum or edges plus nodes. 

\paragraph{Planar trivalent graphs.} Planar graphs are graphs which can be globally  embedded in the plane. 

\begin{definition}
A graph $G \in GRAPH$ is a planar graph, notation $G \in P-GRAPH$, if, forgetting the decoration of the nodes, but not those of the leaves, it can be globally embedded into a disk in the plane such that all in or out leaves are on the boundary of the disk. Two embeddings are considered the same up to  smooth isotopy of disks, which transforms a boundary of a disk into the boundary of the other disk and in or out nodes onto in or out nodes respectively. 
\label{defplan}
\end{definition}

The condition $G \in P-GRAPH$ is global. 

\begin{proposition}
Any graph $G \in GRAPH$ can be transformed into a graph $G' \in P-GRAPH$ by a finite number of 
graphical ($\beta$) moves. 
\label{pplan}
\end{proposition}

\paragraph{Proof.} We may certainly represent the graph  $G \in GRAPH$ into a disk in the plane, such that locally any node and surrounding edges respect the local planar embedding, such that all edges are straight lines  and such that the in and out leaves are on the boundary of the disk. If this is a global embedding then there is nothing to prove. Else, there is a finite number of crossings of the embedded edges. For each pair of embedded edges we eliminate the crossing by an application of a graphical ($\beta$) move and we are done. \quad $\square$

\section{Conversion of lambda terms into $GRAPH$}
\label{constru}

Here we show how to associate to a lambda term a graph in $GRAPH$, then we use this to show that $\beta$-reduction in lambda calculus transforms into the $\beta$ rule for $GRAPH$. 

Indeed, to any term $A \in T(X)$ (where $T(X)$ is the set of lambda terms over the variable 
set $X$) we associate its syntactic tree. The syntactic tree of any lambda term is constructed by using two gates, one corresponding to the $\lambda$ abstraction and the other corresponding to the application. We draw syntactic trees with the leaves (elements of $X$) at the bottom and the root at the top. We shall use the following notation for the two gates: at the left is the gate for the $\lambda$ abstraction and at the right is the gate for the application. 

\vspace{.5cm}

\centerline{\includegraphics[width=60mm]{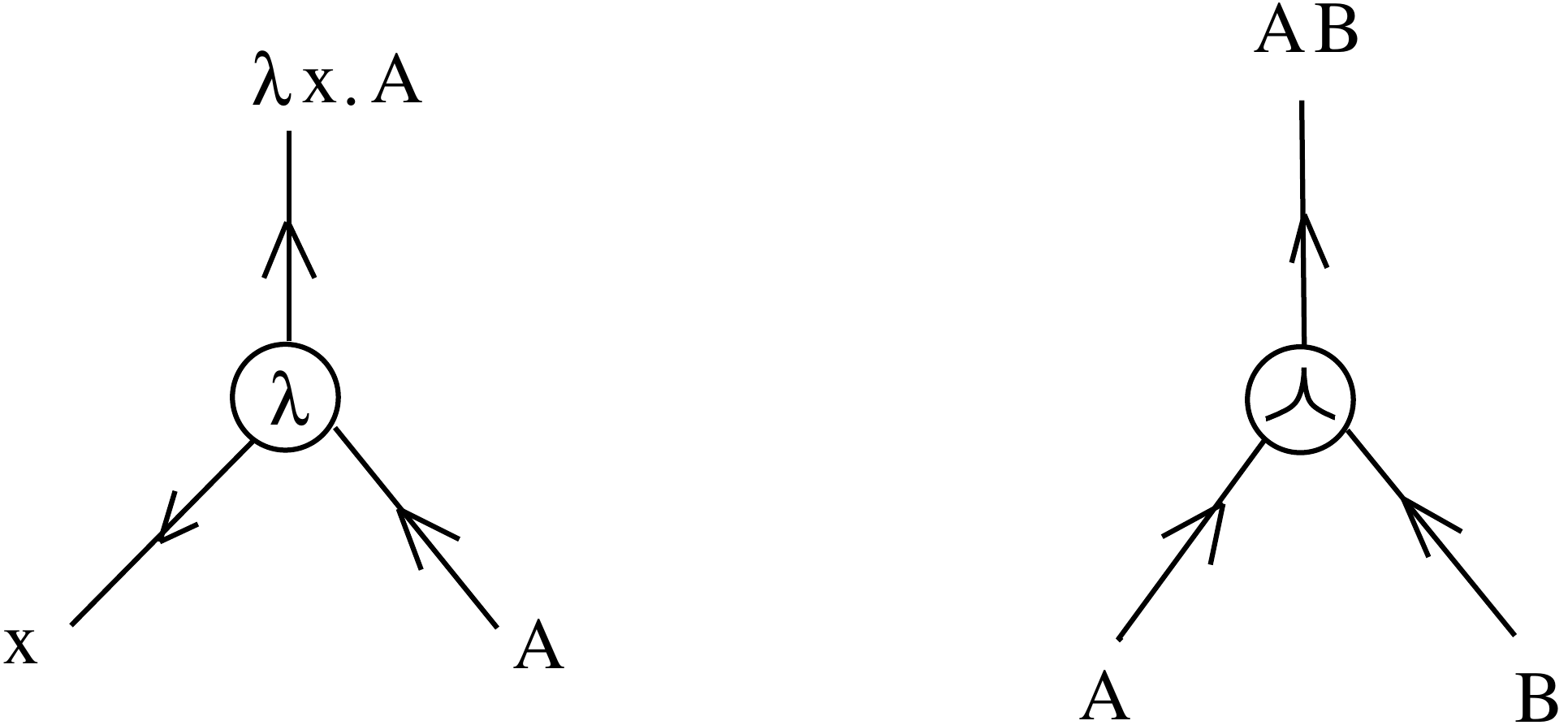}}

\vspace{.5cm}

Remark that these two gates are from the graphical alphabet of $GRAPH$, but the syntactic tree is decorated: at the bottom we have leaves from $X$. Also, remark the peculiar orientation of the edge from the left (in tree notation convention) of the $\lambda$ gate. For the moment, this orientation is in contradiction with the implicit orientation (from down-up) of edges of the syntactic tree, but soon this matter will become clear.

We shall remove all leaves decorations, with the price of introducing new gates, namely $\Upsilon$ and $\top$ gates. This will be done in a sequence of steps, detailed further. Take the syntactic tree of $A \in T(X)$, drawn with the mentioned conventions (concerning gates and the positioning of leaves and root respectively). 

We take as examples the following five lambda terms: $\displaystyle I = \lambda x . x$, 
$\displaystyle  K = \lambda x . (\lambda y. (xy))$, $\displaystyle S = \lambda x . ( \lambda y . 
(\lambda z . ((xz)(yz))))$, $\displaystyle \Omega = (\lambda x. (xx)) (\lambda x. (xx))$ and 
$\displaystyle T =  (\lambda x. (xy)) (\lambda x. (xy))$. 

\paragraph{Step 1.} Elimination of bound variables, part I. Any leaf of the tree is connected to the root by an unique path. 

Start from the leftmost leaf, perform the algorithm explained further, then   go to the right and repeat until all leaves are exhausted. We initialize also a list $B  = \emptyset$ of bound variables.

Take a leaf, say decorated with $x \in X$. To this leaf is associated a word (a list) which is formed by the symbols of gates which are on the path which connects (from the bottom-up) the leaf with the root, together with information about which way, left (L) or right (R), the path passes through the gates. Such a word is formed by the letters $\displaystyle \lambda^{L}$,  
$\displaystyle \lambda^{R}$,  $\displaystyle \curlywedge^{L}$, $\displaystyle \curlywedge^{R}$.

If the first letter is $\displaystyle \lambda^{L}$ then add to the list  $B$ the pair 
$(x, w(x))$ formed by the variable name $x$, and the associated word (describing the path to follow from the respective leaf to the root). Then pass to a new leaf. 

Else continue along the path to the roof. If we arrive at a $\lambda$ gate, this can happen only coming from the right leg of the $\lambda$ gate, thus we can find only the letter $\displaystyle \lambda^{R}$. In such a case look at the variable $y$ which decorates the left leg of the same $\lambda$ gate. If $x =y$ then add to the syntactic tree a new edge, from $y$ to $x$ and proceed further along the path, else proceed further. If the root is attained then pass to next leaf.  

Examples: the graphs associated to the mentioned lambda terms,  together with the list of bound variables, are the following. 

\begin{enumerate}
\item[-] $\displaystyle I = \lambda x . x$ has $\displaystyle B = \left\{ (x, \lambda^{L}) \right\}$, $\displaystyle  K = \lambda x . (\lambda y. (xy))$ has $\displaystyle B = \left\{ (x,  \lambda^{L}), (y, \lambda^{L} \lambda^{R})  \right\}$, $\displaystyle S = \lambda x . ( \lambda y . (\lambda z . ((xz)(yz))))$ has $\displaystyle B = \left\{ (x,  \lambda^{L}), (y, \lambda^{L} \lambda^{R}), (z, \lambda^{L} \lambda^{R} \lambda^{R})  \right\}$. 
\vspace{.5cm}

\centerline{\includegraphics[width=90mm]{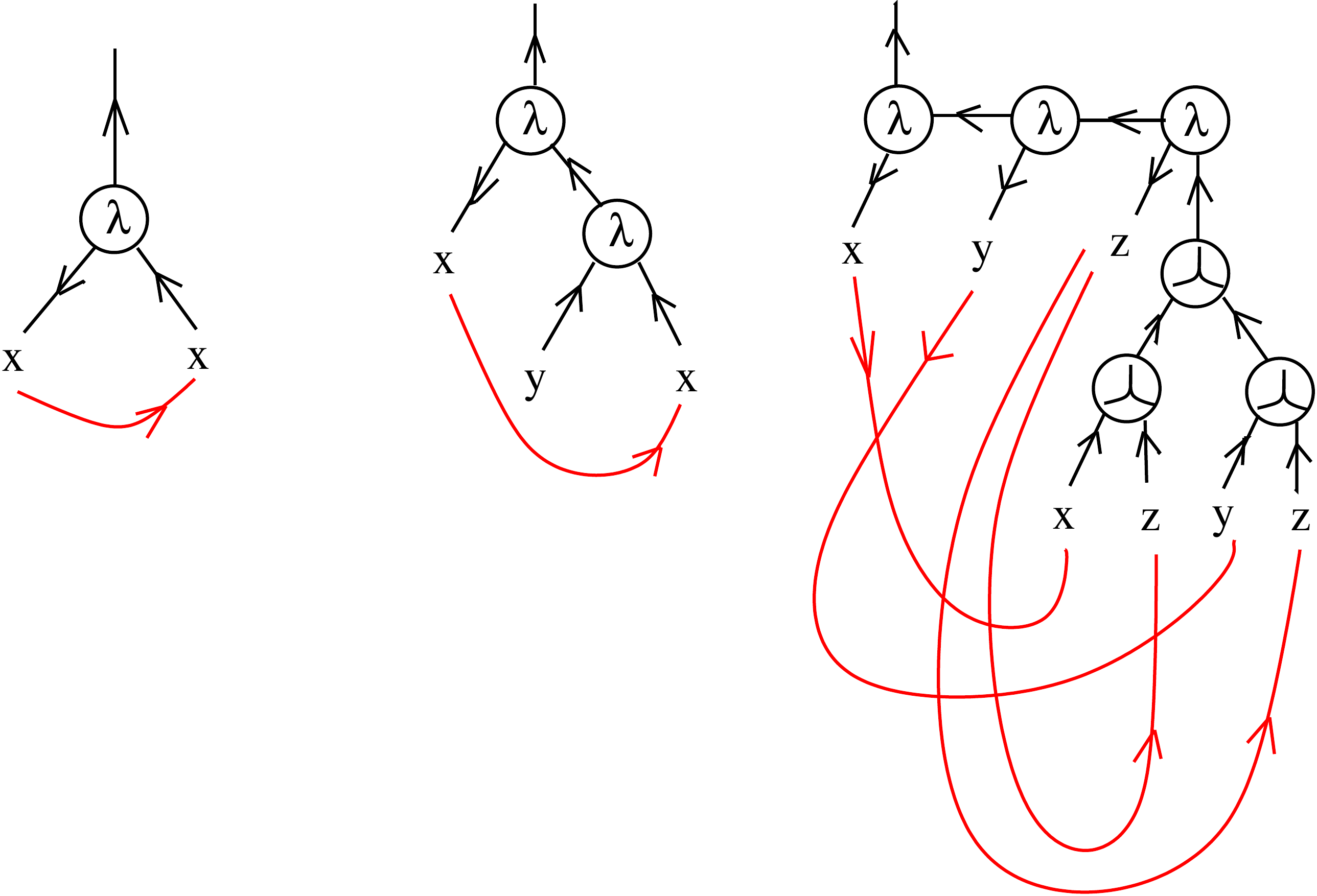}}

\vspace{.5cm}

\item[-]  $\displaystyle \Omega = (\lambda x. (xx)) (\lambda x. (xx))$ has $\displaystyle B = \left\{ (x, \lambda^{L} \curlywedge^{L}) , (x, \lambda^{L} \curlywedge^{R}) \right\}$, 
$\displaystyle T =  (\lambda x. (xy)) (\lambda x. (xy))$ has $\displaystyle B = \left\{ (x, \lambda^{L} \curlywedge^{L}) , (x, \lambda^{L} \curlywedge^{R}) \right\}$. 
\vspace{.5cm}

\centerline{\includegraphics[width=90mm]{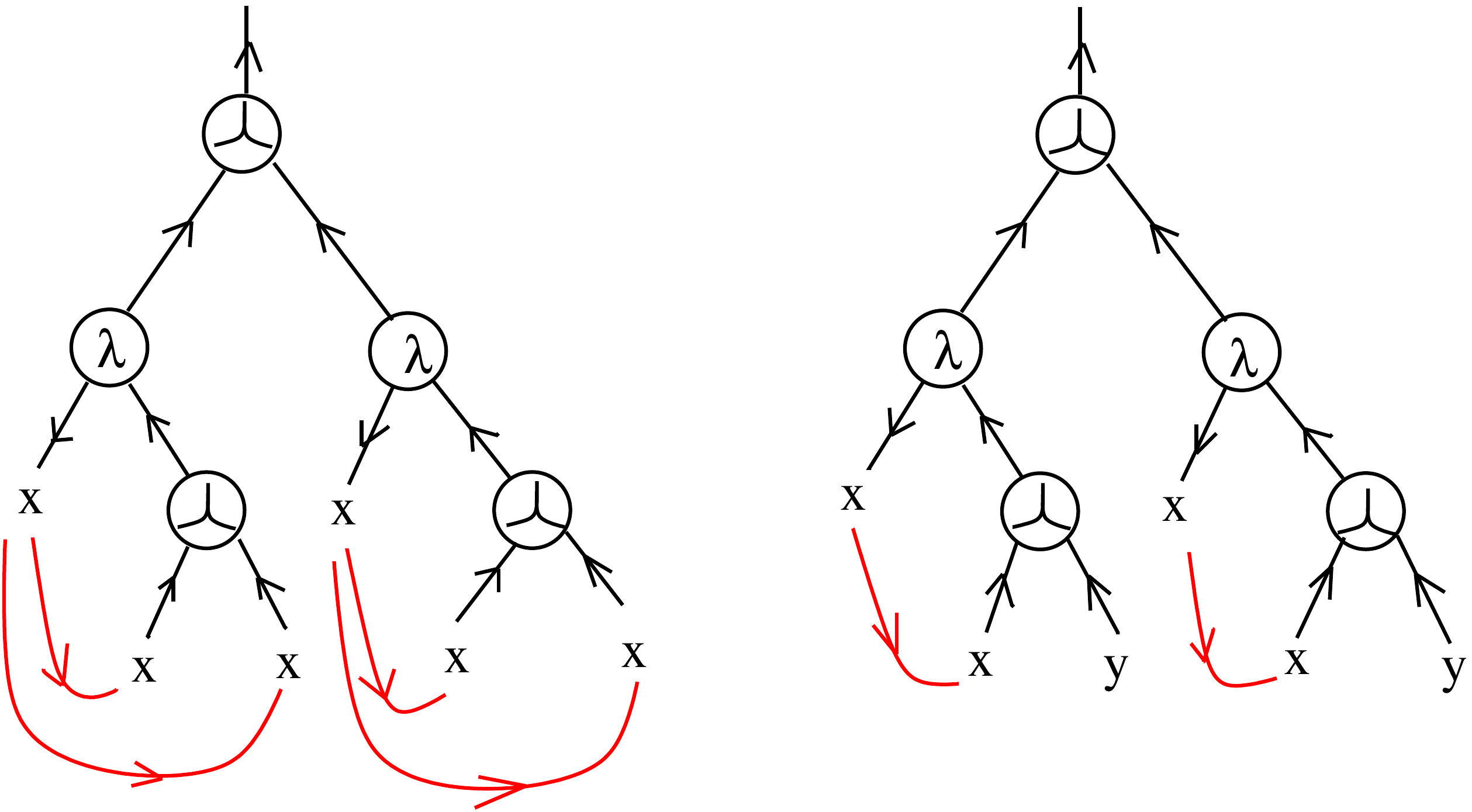}}

\vspace{.5cm}

\end{enumerate}

\paragraph{Step 2.} Elimination of bound variables, part II. We have now a list $B$ of bound variables. If the list is empty then go to the next step. Else, do the following, starting from the first element of the list, until the list is finished. 

An element, say $(x, w(x))$, of the list, is either connected to other leaves by one or more edges added at step 1, or not. If is not connected then erase the variable name with the associated path $w(x)$  and replace it by a $\top$ gate. 
If it is connected then erase it, replace it by a tree formed by $\Upsilon$ gates, which starts at the place where the element of the list were before the erasure and stops at the leaves which were connected to $x$. Erase all decorations which were joined to $x$ and also erase all edges which were added at step 1 to the leave $x$ from the list.

Examples: after the step 2, the graphs associated to the mentioned lambda terms  are the following. 

\begin{enumerate}
\item[-] the graphs of  $\displaystyle I = \lambda x . x$, $\displaystyle  K = \lambda x . (\lambda y. (xy))$, $\displaystyle S = \lambda x . ( \lambda y . (\lambda z . ((xz)(yz))))$ are  
\vspace{.5cm}

\centerline{\includegraphics[width=90mm]{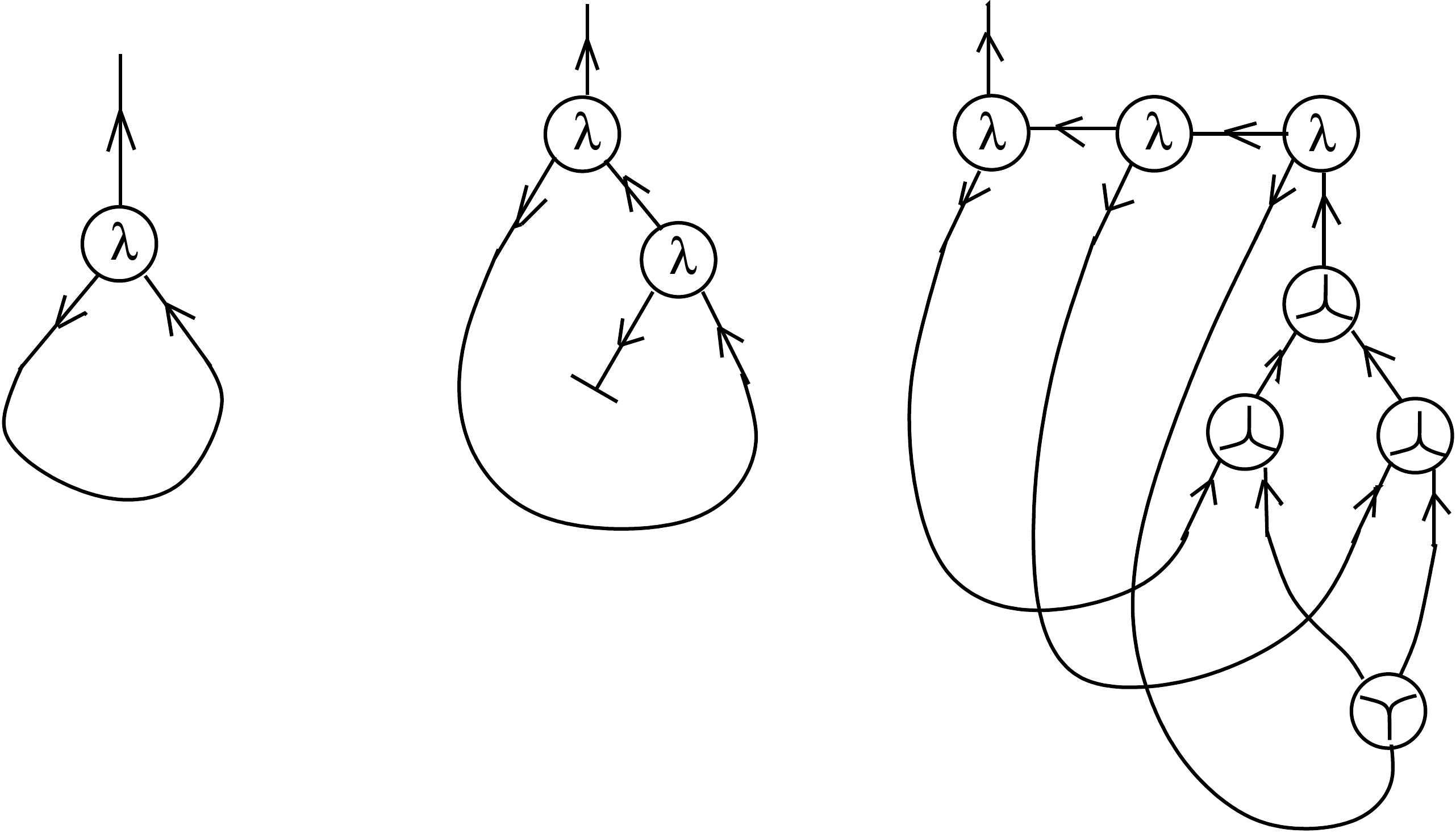}}

\vspace{.5cm}

\item[-]  the graphs of $\displaystyle \Omega = (\lambda x. (xx)) (\lambda x. (xx))$,  
$\displaystyle T =  (\lambda x. (xy)) (\lambda x. (xy))$ are  
\vspace{.5cm}

\centerline{\includegraphics[width=90mm]{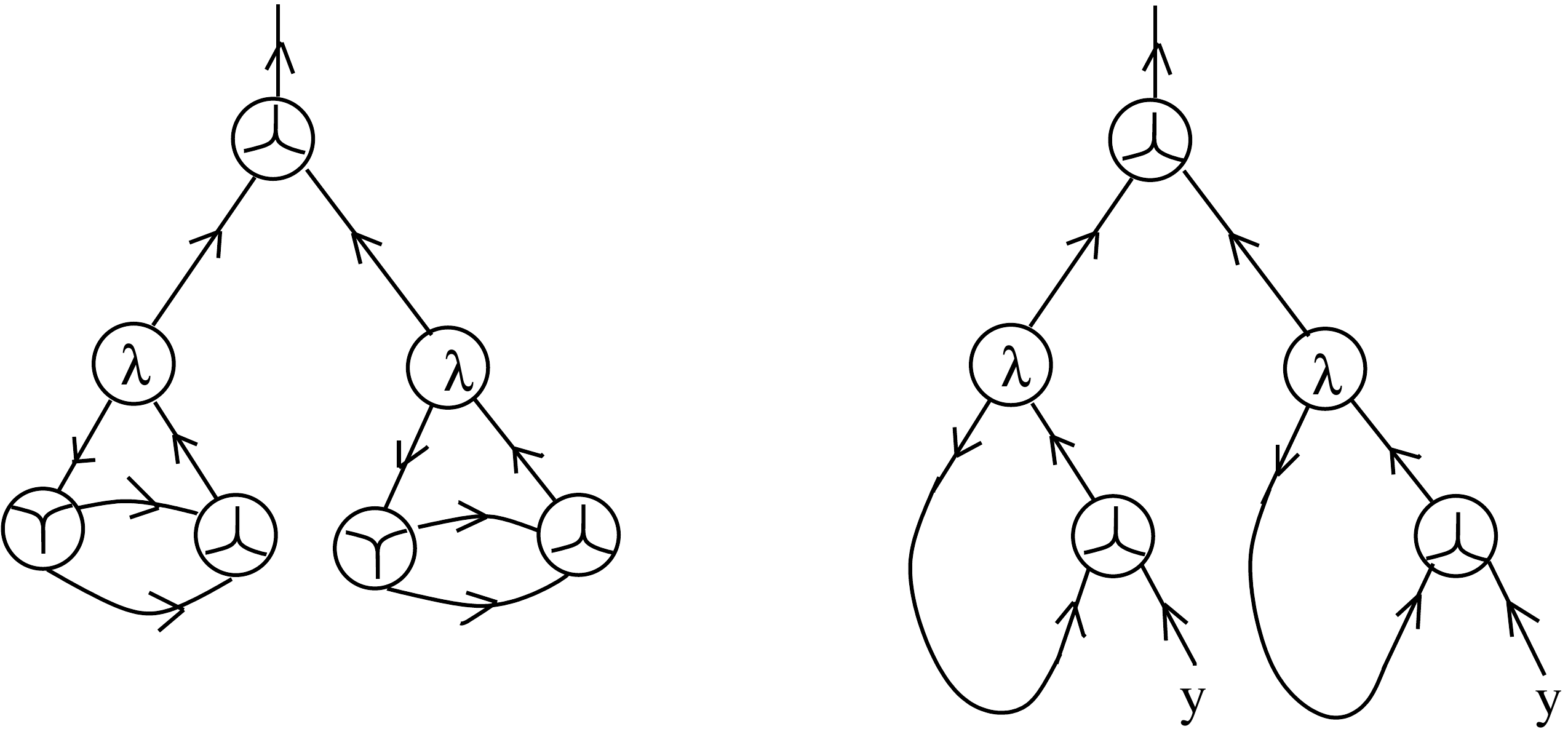}}

\vspace{.5cm}

\end{enumerate}

Remark that at this step the necessity of having the peculiar orientation of the left leg of the $\lambda$ gate becomes clear. 

Remark also that there may be more than one possible tree of gates $\Upsilon$, at each elimination of a bound variable (in case a bound variable has at least tree occurrences). One may use any tree of $\Upsilon$ which is fit. The problem of multiple possibilities is the reason of introducing the (CO-ASSOC) move.  

\paragraph{Step 3.} We may still have leaves decorated by free variables. Starting from the left to the right, group them together in case some of them occur in multiple places, then replace the multiple occurrences of a free variable by a tree of $\Upsilon$ gates with a free  root, which ends exactly where the occurrences of the respective variable are. Again, there are multiple ways of doing this, but we may pass from one to another by a sequence of (CO-ASSOC) moves. 

Examples: after the step 3,  all the graphs associated to the mentioned lambda terms, excepting the last one, are left unchanged. The graph of the last term, changes. 

\begin{enumerate}
\item[-]  as an illustration, we figure the graphs of $\displaystyle \Omega = (\lambda x. (xx)) (\lambda x. (xx))$,  left unchanged by step 3, and the graph of 
$\displaystyle T =  (\lambda x. (xy)) (\lambda x. (xy))$:
\vspace{.5cm}

\centerline{\includegraphics[width=90mm]{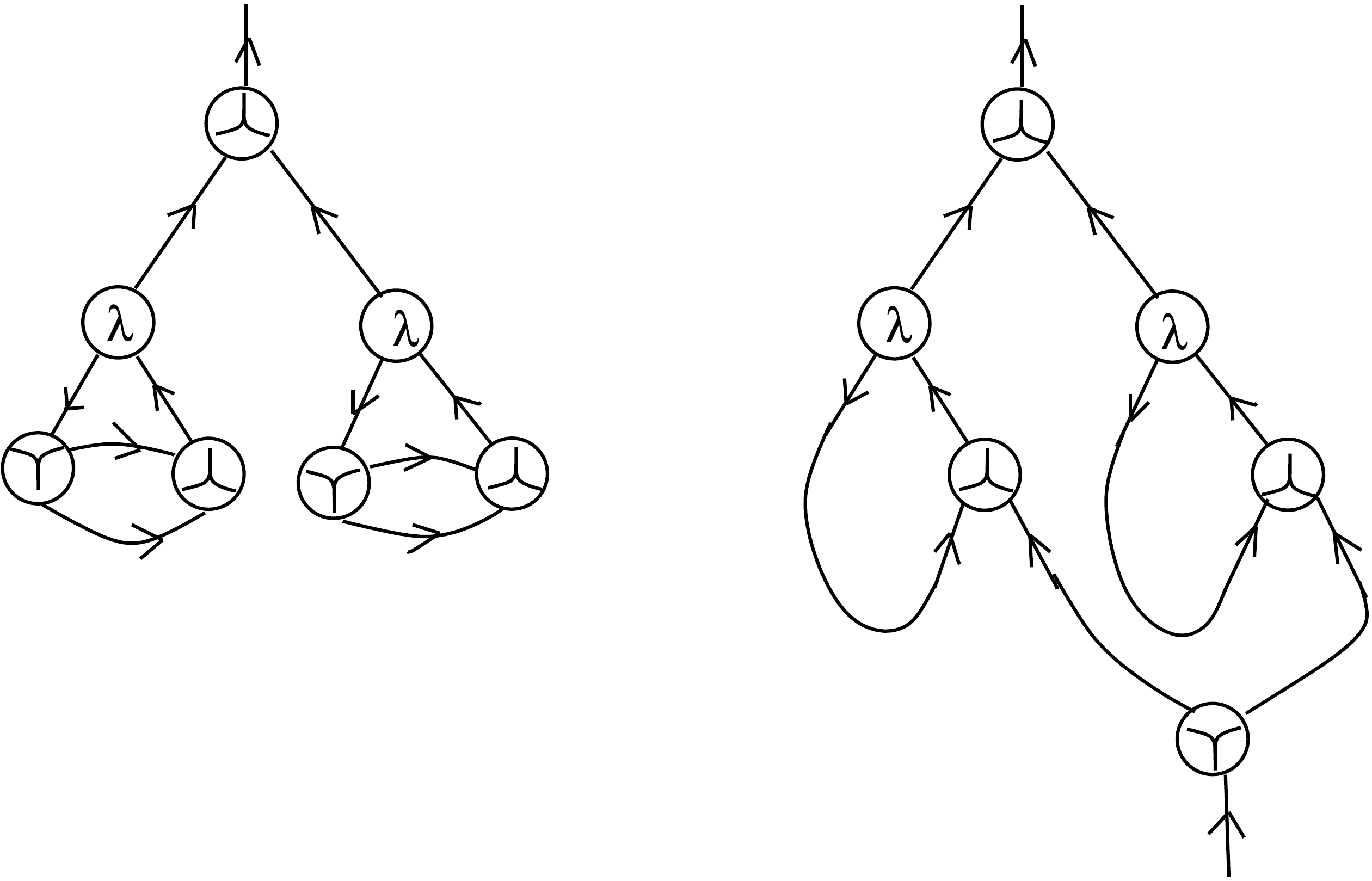}}

\vspace{.5cm}

\end{enumerate}

\begin{theorem}
Let $A \mapsto [A]$ be a transformation of a lambda term $A$ into a graph $[A]$ as described previously (multiple transformations are possible because of the choice of $\Upsilon$ trees). Then: 
\begin{enumerate}
\item[(a)] for any term $A$ the graph $[A]$ is in $\lambda GRAPH$, 
\item[(b)] if $[A]'$ and $[A]"$ are transformations of the term $A$ then we may pass from 
$[A]'$ to $[A]"$ by using a finite number (exponential in the number of leaves of the syntactic tree of $A$) of (CO-ASSOC) moves,
\item[(c)] if $B$ is obtained from $A$ by $\alpha$-conversion then we may pass from  $[A]$ to $[B]$ by a finite sequence of (CO-ASSOC) moves, 
\item[(d)] let $A, B \in T(X)$ be two terms and $x \in X$ be a variable. Consider the terms 
$\lambda x . A$ and $A[x:=B]$, where $A[x:=B]$ is the term obtained by substituting in $A$ the free occurrences of $x$ by $B$. We know that $\beta$ reduction in lambda calculus consists in passing from $(\lambda x . A) B$ to $A[x:=B]$.  Then, by one $\beta$ move in $GRAPH$ applied to  $[(\lambda x . A) B]$ we pass to a graph which can be further  transformed into one  of  $A[x:=B]$, via (global FAN-OUT) moves,  (CO-ASSOC) moves and pruning moves. 
\end{enumerate}
\label{lambdathm}
\end{theorem} 

\paragraph{Proof.} (a) we have to prove that for any node $\lambda$ any oriented path in $[A]$ starting at the left exiting edge  of this node can be completed to a path which either terminates in a graph $\top$, or else terminates at the entry peg of this node, but this is clear. Indeed, either the bound variable (of this $\lambda$ node in the syntactic tree of $A$) is fresh, then the bound variable is replaced by a $\top$ gate, or else, the bound variable is replaced by a tree of $\Upsilon$ gates. No matter which path we choose, we may complete it to a cycle passing by the said $\lambda$ node.

(b) Clear also, because the (CO-ASSOC) move is designed for passing from a tree of $\Upsilon$ gates to another tree with the same number of leaves. 

(c) Indeed, the names of bound variables of $A$ do not affect the construction of $[A]$, therefore if $B$ is obtained by $\alpha$-conversion of $A$, then $[B]$ differs from $[A]$ only by the particular choice of  trees of $\Upsilon$ gates. But this is solved by (CO-ASSOC) moves. 

(d) This is the surprising, maybe, part of the theorem. There are two cases: $x$ is fresh for $A$ or not. If $x$ is fresh for $A$ then in the graph $[(\lambda x.A) B]$ the name variable $x$ is replaced by a $\top$ gate. If not, then all the occurrences of $x$ in $A$ are connected by a $\Upsilon$ tree with root at the left peg of the $\lambda$ gate where $x$ appears as a bound variable.

In the case when $x$ is not fresh for $A$, we see in the LHS of the figure the graph  $[(\lambda x . A) B]$  (with a remanent decoration of "x").  We perform a graphic ($\beta$) move and we obtain the graph from the right.

\vspace{.5cm}

\centerline{\includegraphics[width=90mm]{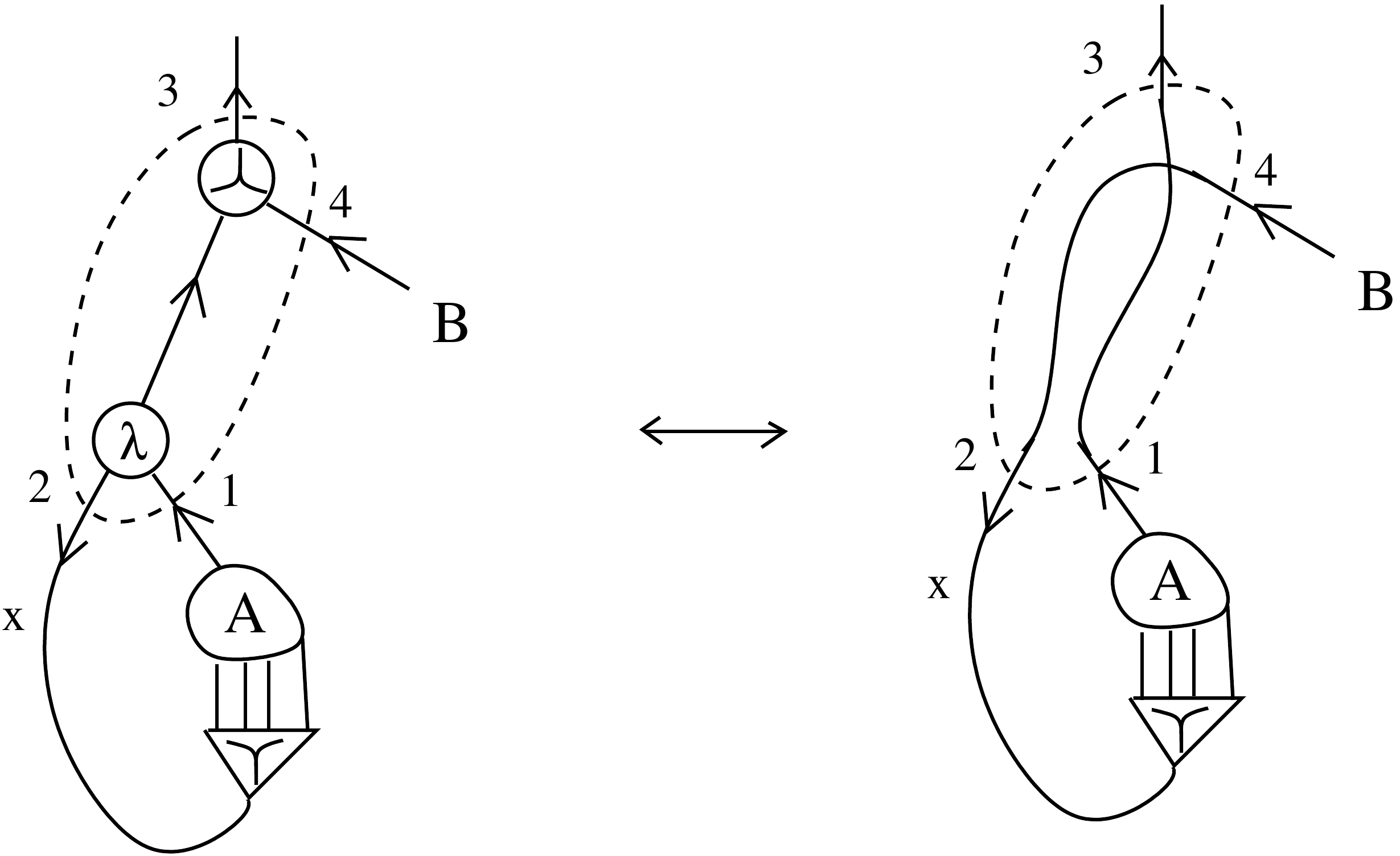}}

\vspace{.5cm}

This graph can be transformed into a graph of $A[x:=B]$ via (global FAN-OUT) and (CO-ASSOC)  moves. 
The case when  $x$ is fresh for $A$ is figured next.

\vspace{.5cm}

\centerline{\includegraphics[width=90mm]{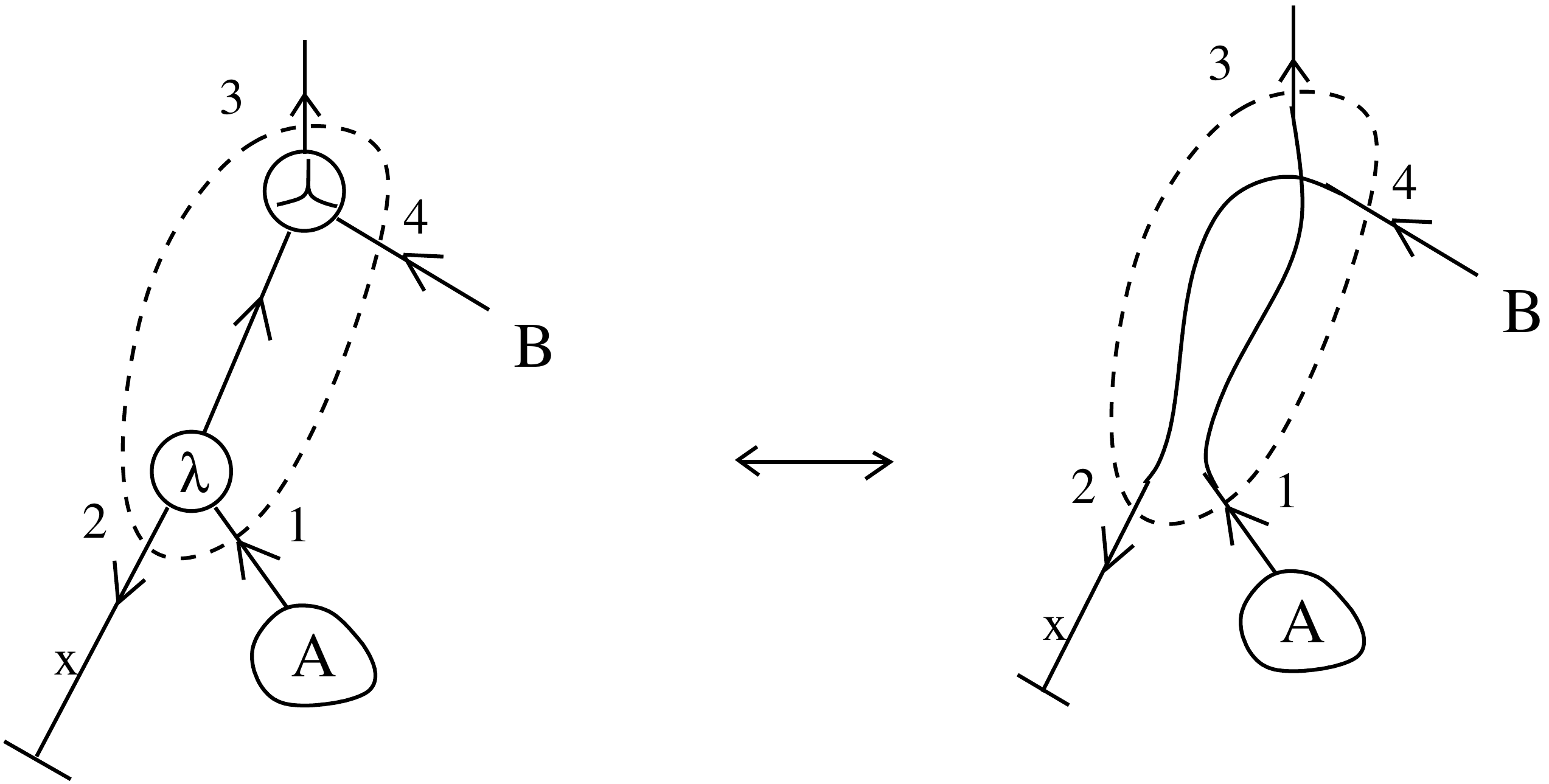}}

\vspace{.5cm}

We see that the graph obtained by performing the graphic ($\beta$) move is the union of the graph of $A$ and the graph of $B$ with a $\top$ gate added at the root. By pruning we are left with the graph of $A$, which is consistent to the fact that when  $x$ is fresh for $A$ then 
$(\lambda x . A) B$ transforms by  $\beta$ reduction into $A$.  \quad $\square$

As an example, let us manipulate the graph of $\displaystyle \Omega = (\lambda x. (xx)) (\lambda x. (xx))$: 

\vspace{.5cm}

\centerline{\includegraphics[width=90mm]{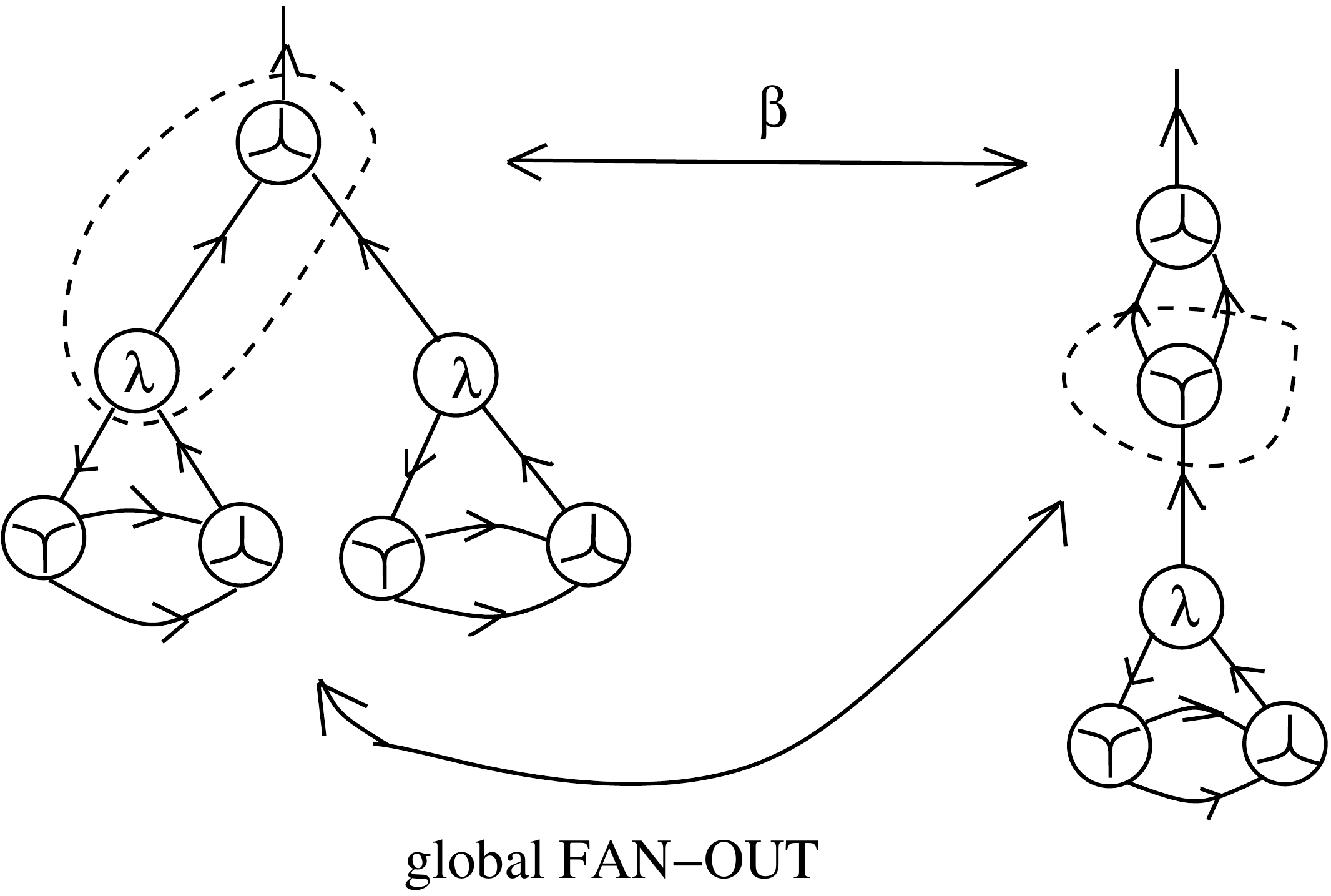}}

\vspace{.5cm}

We can pass from the LHS figure to the RHS figure by using a graphic ($\beta$) move. Conversely, we can pass from the RHS figure to the LHS figure by using a (global FAN-OUT) move. 
These manipulations correspond to the well known fact that $\Omega$ is left unchanged after $\beta$ reduction:  let $\displaystyle U = \lambda x. (xx)$, then 
$\displaystyle \Omega = U U = (\lambda x. (xx)) U \leftrightarrow  U U = \Omega$.

\subsection{Examples: combinatory logic and arithmetic}

\paragraph{$S$, $K$ and $I$ combinators in $GRAPH$.} The combinators $\displaystyle I = \lambda x . x$, $\displaystyle  K = \lambda x . (\lambda y. (xy))$ and $\displaystyle S = \lambda x . ( \lambda y . (\lambda z . ((xz)(yz))))$ have the following correspondents in $GRAPH$, denoted by the same letters: 

\vspace{.5cm}

\centerline{\includegraphics[width=100mm]{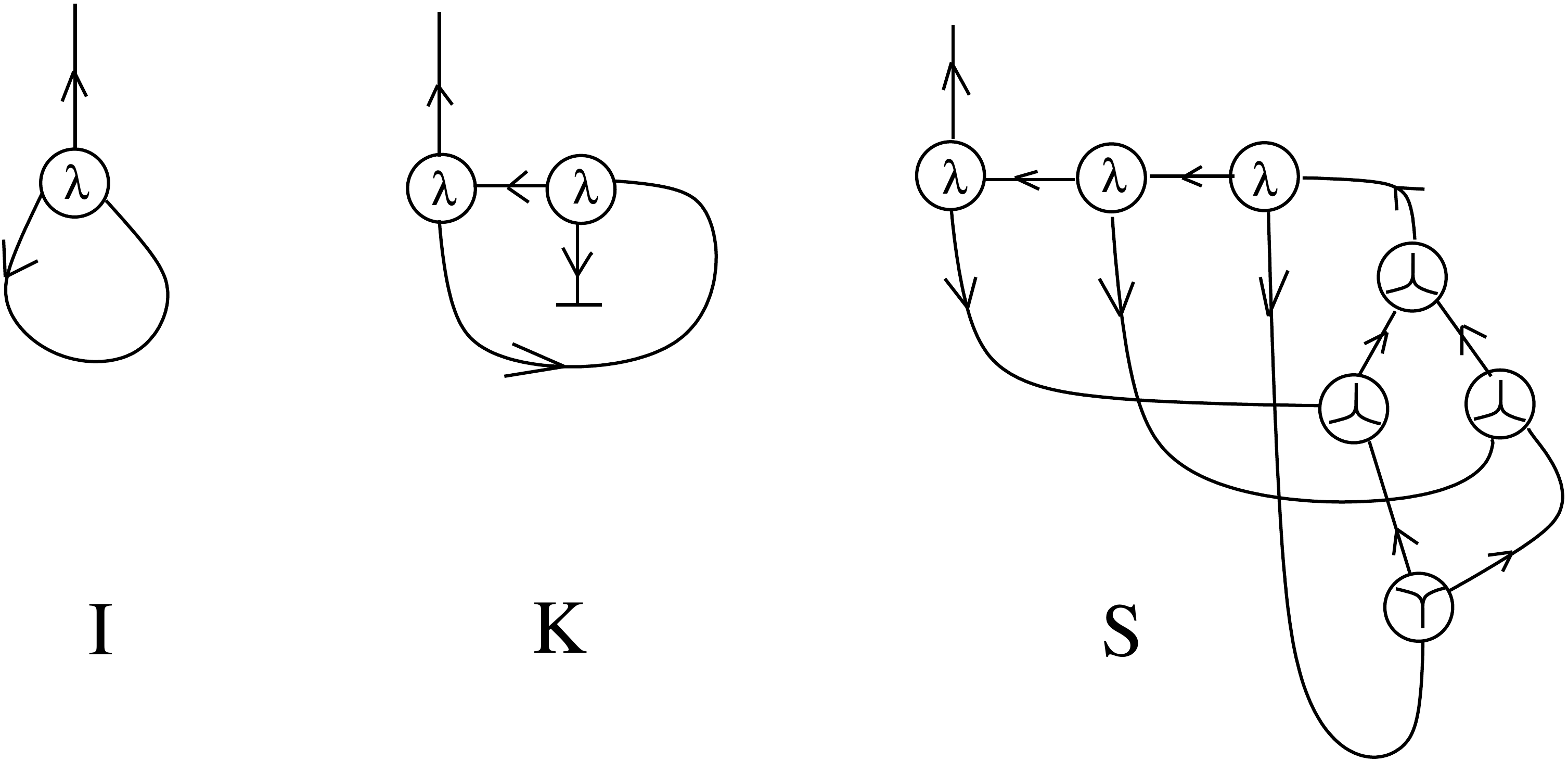}}

\vspace{.5cm}

\begin{proposition}
(a) By one graphic ($\beta$) move $I \curlywedge A$ transforms into $A$, for any $A \in GRAPH$ with one output. 

(b) By two graphic ($\beta$) moves, followed by a global pruning, for any $A, B \in GRAPH$ with one output, the graph $(K \curlywedge A) \curlywedge B$ transforms into $A$. 

(c) By five graphic ($\beta$) moves, followed by one local pruning move, the graph $(S \curlywedge K) \curlywedge K$ transforms into $I$. 

(d) By three graphic ($\beta$) moves followed by a (global FAN-OUT) move, for any  $A, B, C \in GRAPH$ with one output, the graph 
$((S \curlywedge A)\curlywedge B) \curlywedge C$ transforms into the graph $(A \curlywedge C) \curlywedge (B \curlywedge C)$. 
\label{pcombi}
\end{proposition}

\paragraph{Proof.} The proof of (b) is given in the next figure. 

\vspace{.5cm}

\centerline{\includegraphics[width=120mm]{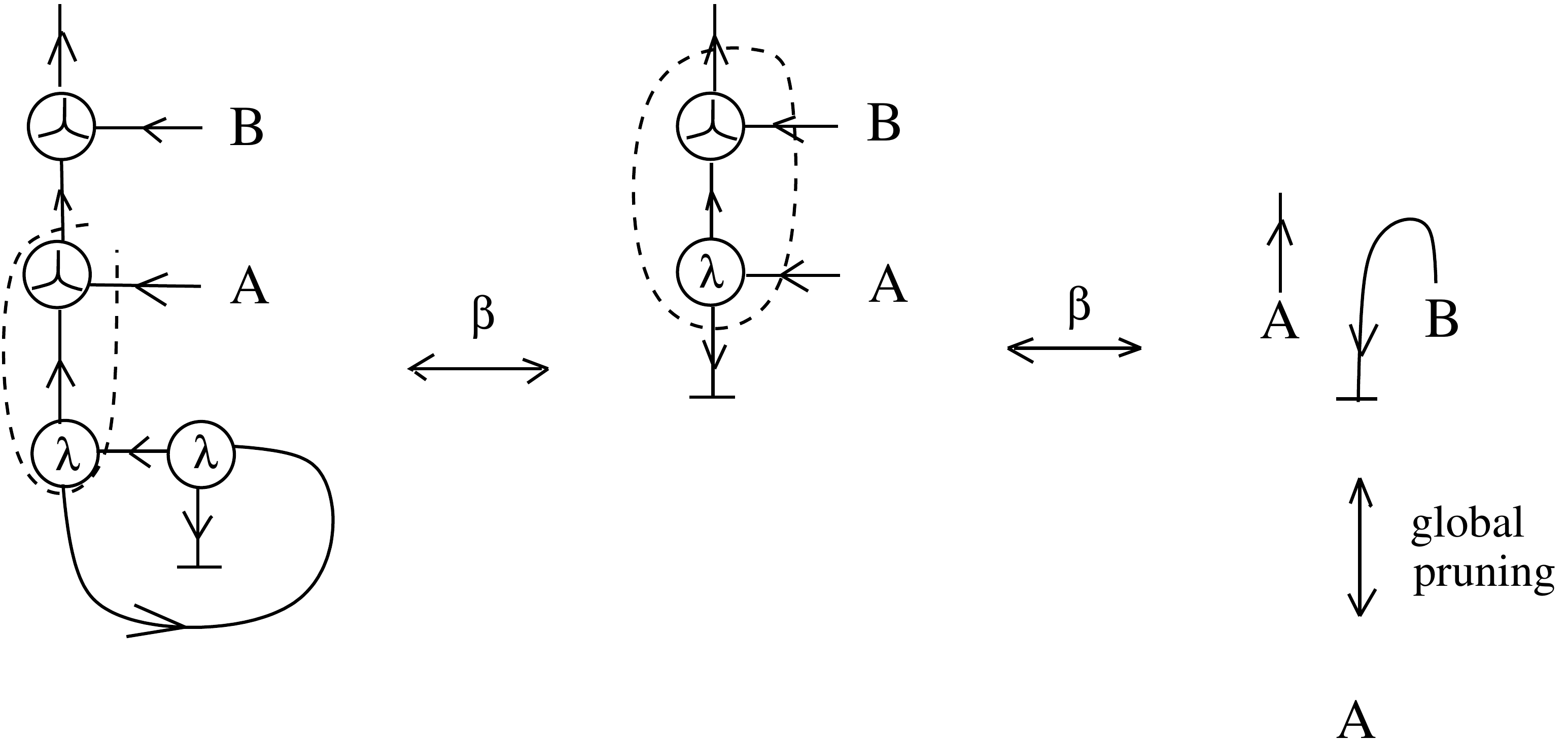}}

\vspace{.5cm}

The proof of (c) is given in the following figure. 

\vspace{.5cm}

\centerline{\includegraphics[width=120mm]{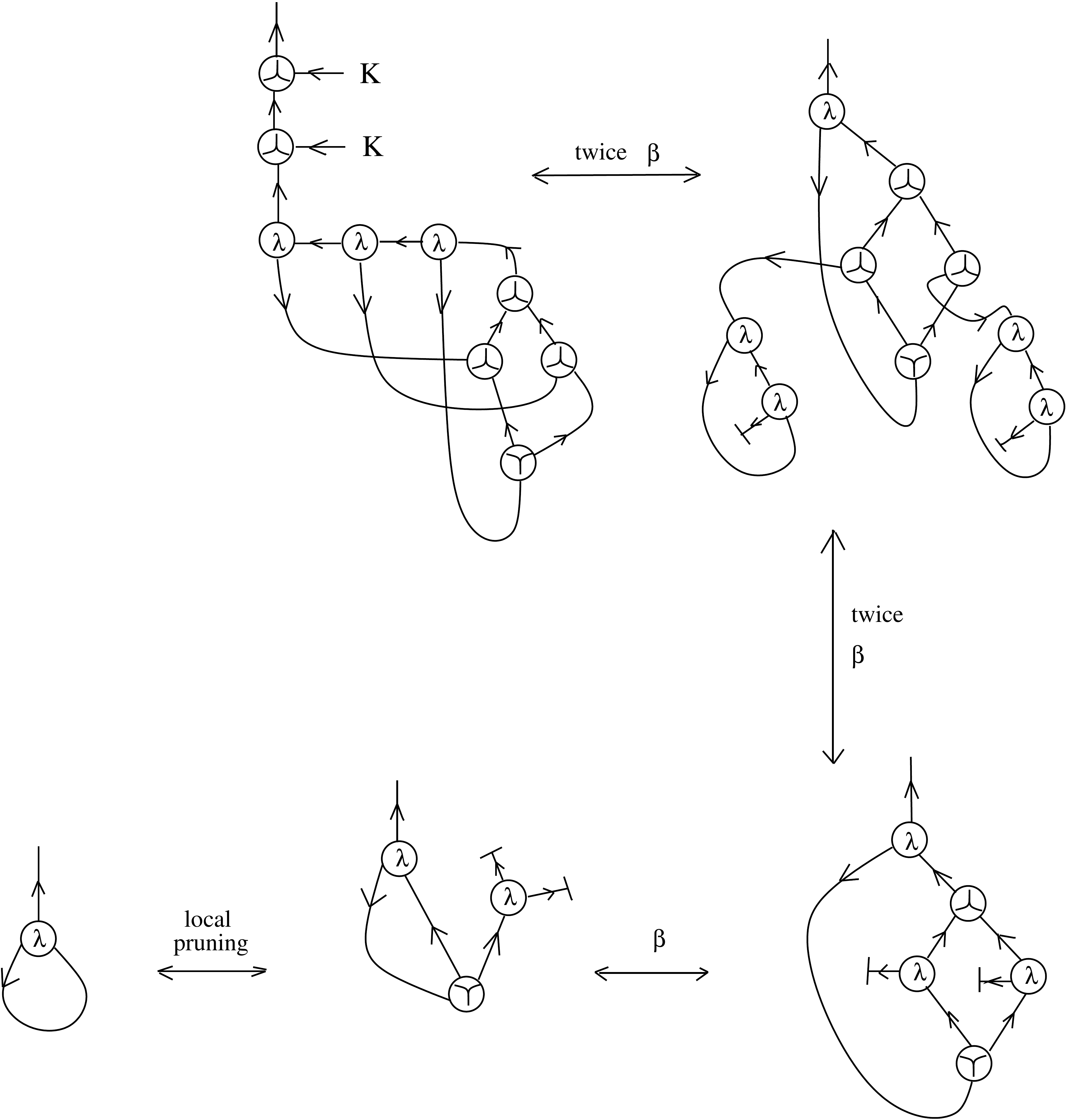}}

\vspace{.5cm}

 (a) and (d) are left to the interested reader. \quad $\square$

\paragraph{Church numerals.} We shall introduce graphs which represent natural numbers and also a graph called successor, constructed from the  Church numerals and the encoding of lambda calculus in $GRAPH$,  according to the procedure described in section \ref{constru}. 

\begin{definition}
The following graphs represent natural numbers: 
\begin{enumerate}
\item[-] the "$0$" graph is:

\vspace{.5cm}

\centerline{\includegraphics[width=15mm]{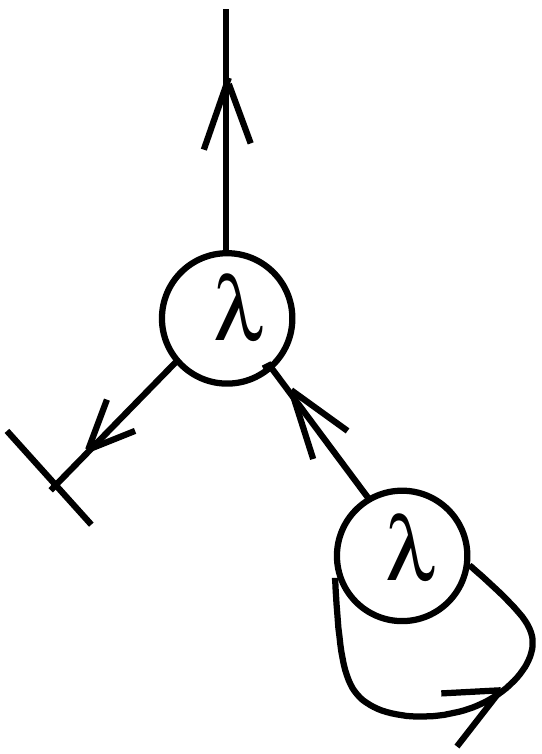}}

\vspace{.5cm}

\item[-] the "$1$", "$2$", "$3$" graphs are, from left to right:

\vspace{.5cm}

\centerline{ \includegraphics[width=100mm]{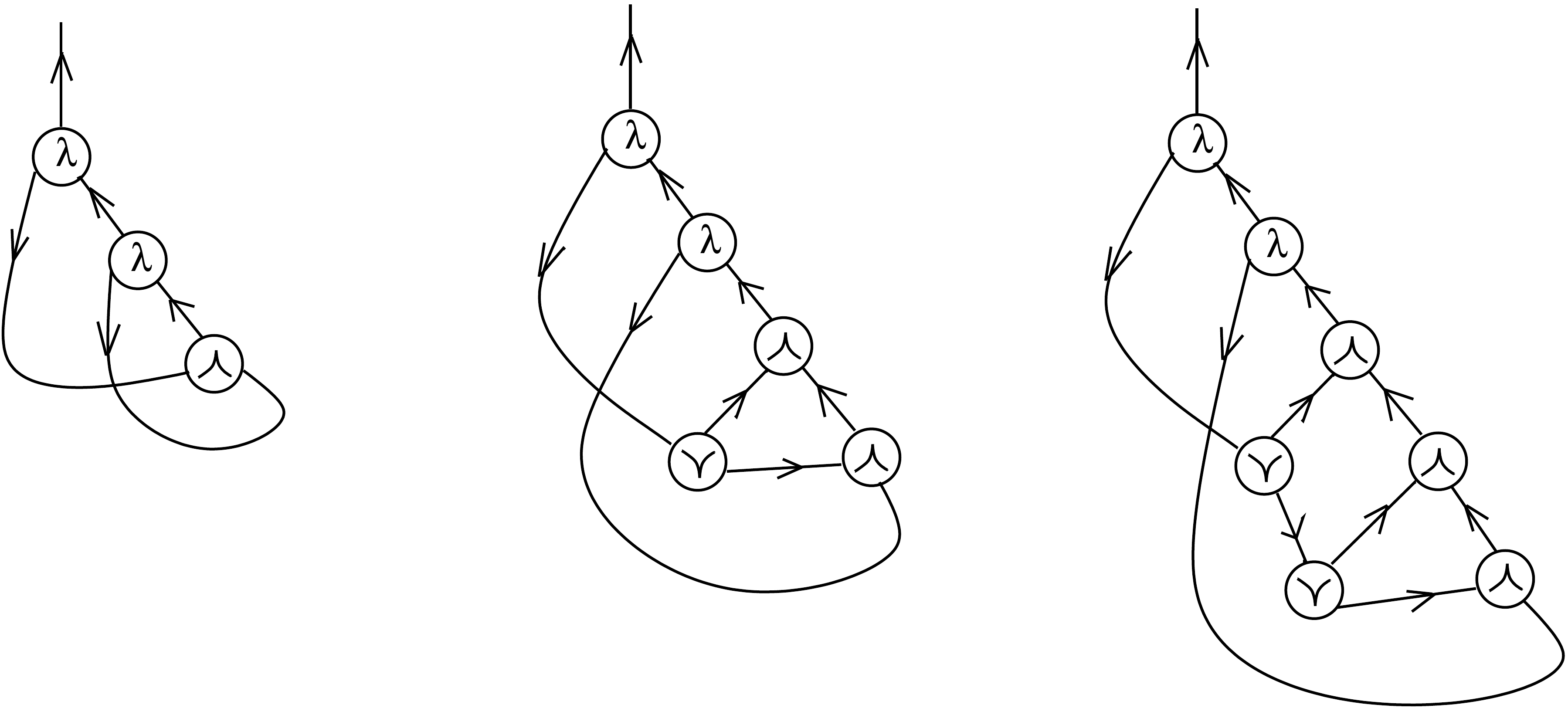}}

\vspace{.5cm}

\item[-] for any natural number $n \geq 3$, the "$n$" graph is:

\vspace{.5cm}

\centerline{ \includegraphics[width=80mm]{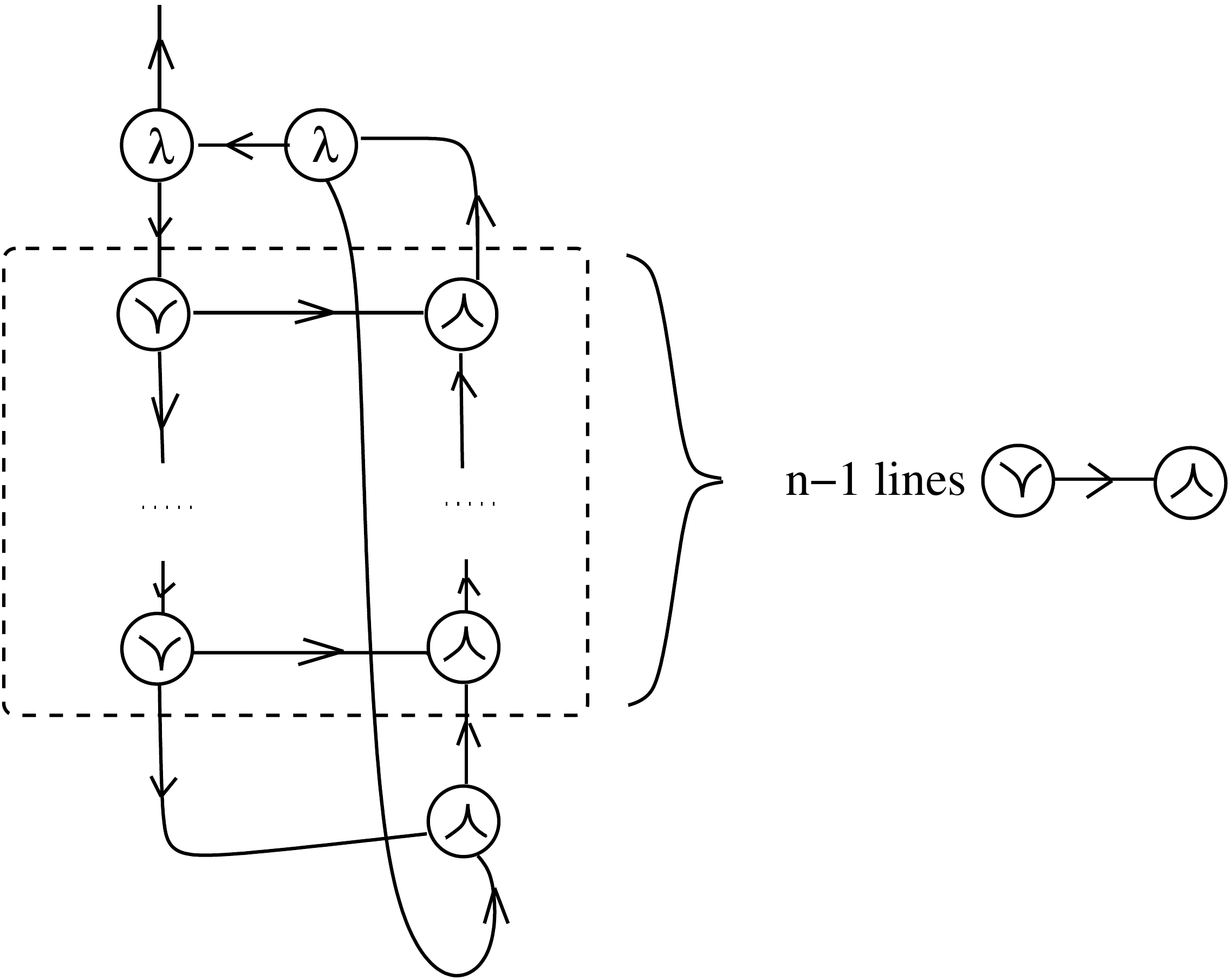}}

\vspace{.5cm}

\item[-] the succesor graph "$SUCC$" is:

\vspace{.5cm}

\centerline{ \includegraphics[width=40mm]{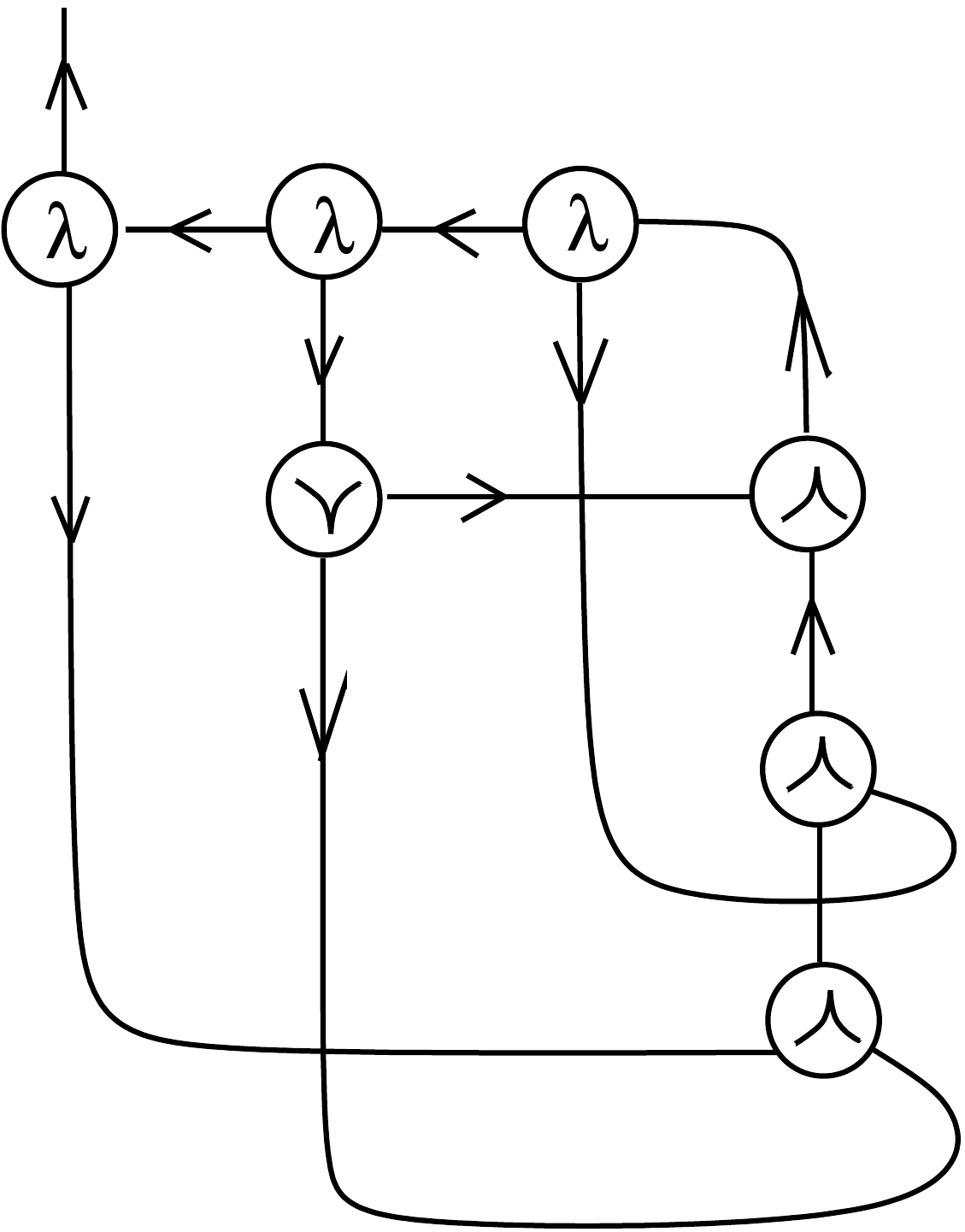}}

\vspace{.5cm}

\item[-] the plus graph "$PLUS$" is:

\vspace{.5cm}

\centerline{ \includegraphics[width=40mm]{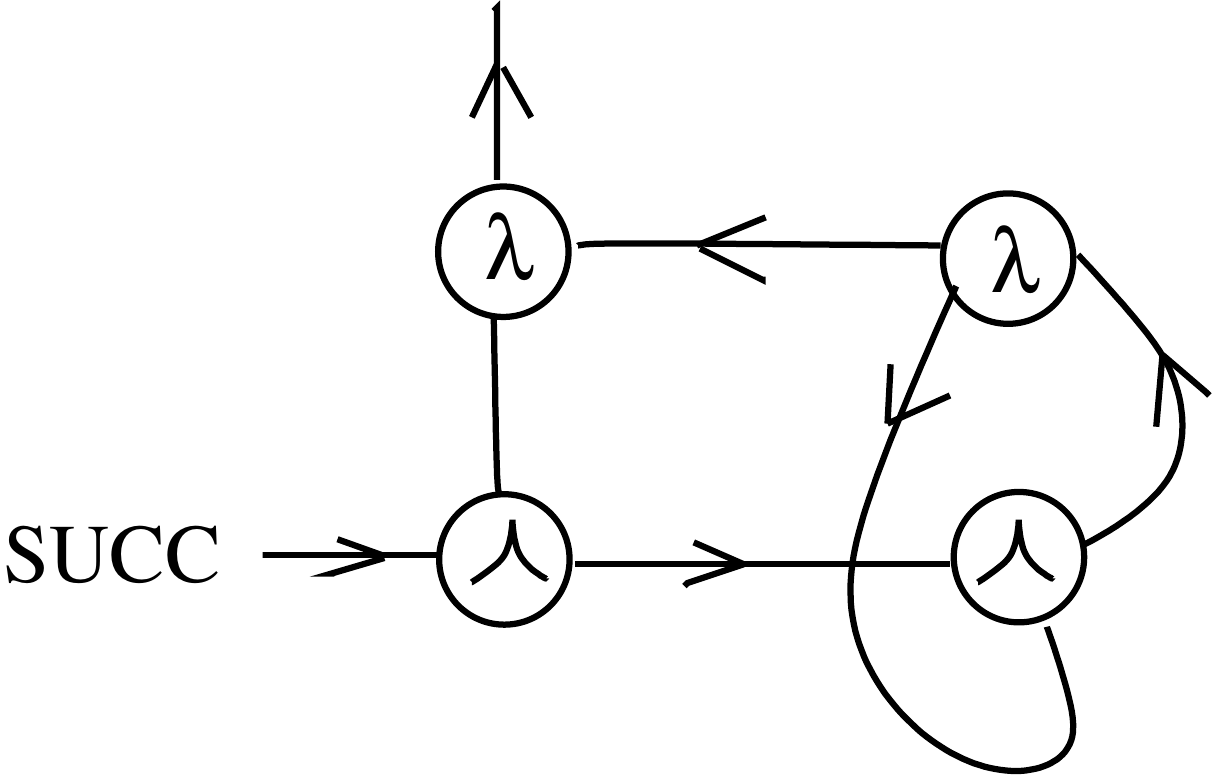}}

\vspace{.5cm}

\end{enumerate}
\end{definition}

\begin{proposition}
By using three ($\beta$) moves, the graph obtained by applying $SUCC$ to $n$, i.e. the graph 
$SUCC \curlywedge n$, is transformed into the graph $n+1$. 
\end{proposition}

\paragraph{Proof.} 
The graph $SUCC \curlywedge n$ is this: 
\vspace{.5cm}

\centerline{ \includegraphics[width=25mm]{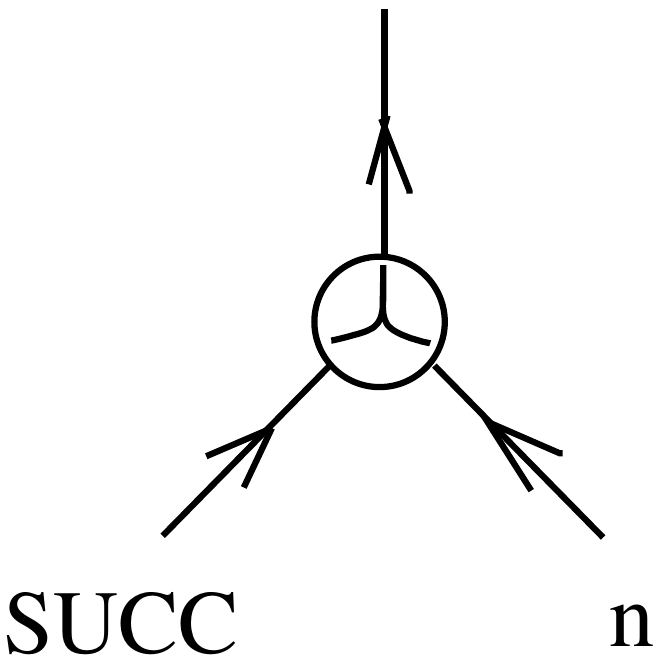}}

\vspace{.5cm}

Starting from this graph, we shall apply three ($\beta$) moves. Each time we shall mark by a closed dashed curve the place where the $\beta$ rule is applied.

We leave to the reader the cases $n=0, 1, 2$ and we give the proof for $n \geq 3$. 

The first ($\beta$) move applies to the graph $SUCC \curlywedge n$, figured in all details, at the place encircled by a curved dashed line: 

\vspace{.5cm}

\centerline{ \includegraphics[width=100mm]{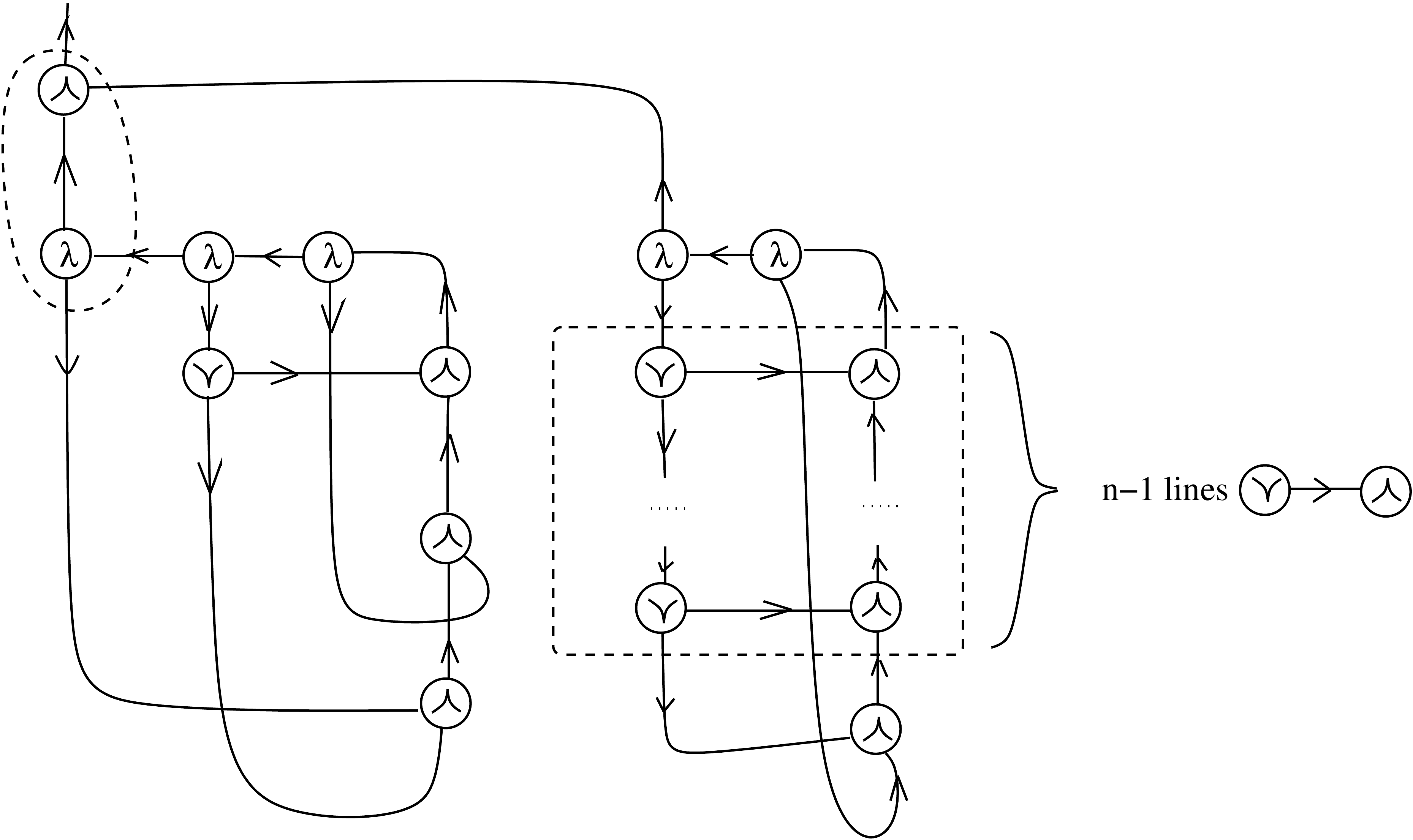}}

\vspace{.5cm}

After the first ($\beta$) move we obtain the graph figured here: 

\vspace{.5cm}

\centerline{ \includegraphics[width=100mm]{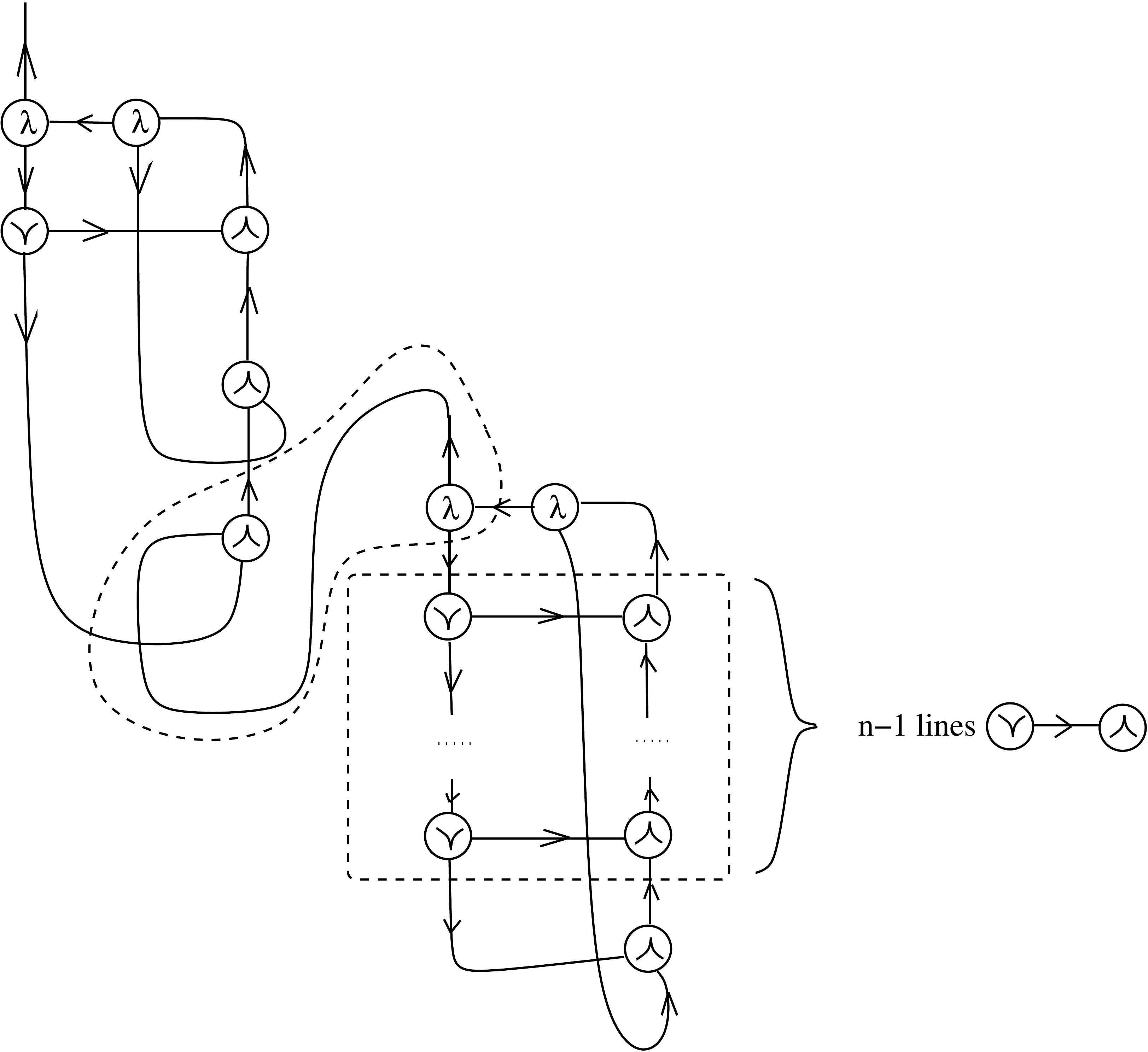}}

\vspace{.5cm}

We see another place encircled by a curved dashed line, where we shall apply the second ($\beta$) move. The result is this: 

\vspace{.5cm}

\centerline{ \includegraphics[width=80mm]{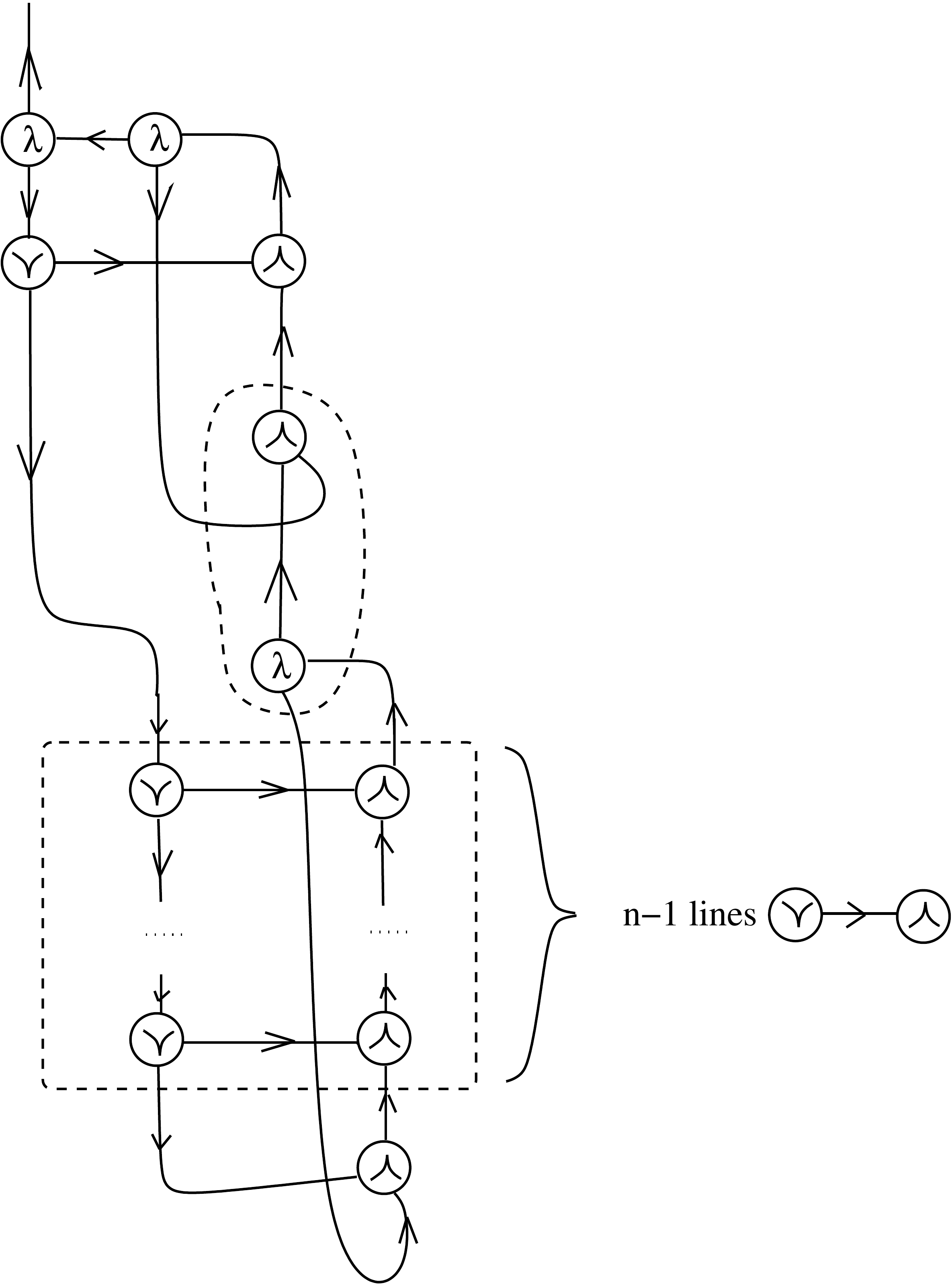}}

\vspace{.5cm}

Finally, we obtain the graph $n+1$ by applying  the third ($\beta$) move at the place indicated  by a curved dashed line. Indeed, remark that after this last ($\beta$) move a new line $\Upsilon \rightarrow \curlywedge$ adds to the $n-1$ previous lines. \quad $\square$

In order for the "plus" operation to work properly, we need to use (global FAN-OUT) moves. 

\begin{proposition}
By using  ($\beta$) moves  and (global FAN-OUT) moves, the graph obtained by  $(PLUS \curlywedge m) \curlywedge n$, can be transformed into the graph $m+n$. 
\end{proposition}

The proof is left to the interested reader. 

\section{$\lambda$-Scale calculus and moves in $GRAPH$} 

For $\lambda$-Scale calculus see \cite{lambdascale}. Terms in this calculus admit a transformation of their respective syntactic trees into graphs in $GRAPH$. The procedure is similar to the one described in section \ref{constru}. The syntactic tree of a term  in $\lambda$-Scale is constructed from the gates $\lambda$, $\curlywedge$ and $\bar{\varepsilon}$ with 
$\varepsilon \in \Gamma$, where $\Gamma$ is a given abelian group. 

\paragraph{$\lambda$-Scale calculus.} In this calculus we have two operations: 
\begin{enumerate}
\item[-] the lambda abstraction, denoted by $(x,A) \mapsto x \lambda A$, with $x$ a variable (name) and $A$ a term, 
\item[-] for any $\varepsilon \in \Gamma$ (where $\Gamma$ is a given abelian group) there is an operation $(A,B) \mapsto A \varepsilon B$, with $A$, $B$ terms. 
\end{enumerate}

The terms, variables, free and bound, $\alpha$-conversion and substitution in $\lambda$-Scale calculus are defined as usual. The moves in  $\lambda$-Scale calculus are: 
\begin{enumerate}
\item[($\beta$*)] if  $y \not \in FV(B) \cup FV(A[x:=B])$ then 
$\displaystyle (x \lambda A) \varepsilon B \leftrightarrow (y \lambda (A[x:=B]))\varepsilon B$
\item[(R1)] if $x \not \in FV(A)$ then for any $\varepsilon \in \Gamma$ 
$(x\lambda A) \varepsilon A \leftrightarrow A$, 
\item[(R2)] if $x \not \in FV(B)$ and $\varepsilon, \mu \in \Gamma$ then 
$(x \lambda (B \mu x)) \varepsilon A \leftrightarrow B (\varepsilon \mu) A$, 
\item[(ext1)] if $x \not \in FV(B)$ then 
$x \lambda (B 1 x) \leftrightarrow B$, 
\item[(ext2)] if $x \not \in  FV(B)$ then $(x \lambda B) 1 A \leftrightarrow B$. 
\end{enumerate}

We proved in theorem 3.5 \cite{lambdascale} that $\lambda$-Scale calculus contains lambda calculus with $\eta$-reduction.

The syntactic tree of a term is made by elementary gates $\lambda$ and $\displaystyle \hat{\varepsilon}$, the former corresponding to the $\lambda$ abstraction and the latter to the $\varepsilon$ operation. For the $\lambda$ elementary tree we adopt the same convention of drawing as before. 

\vspace{.5cm}

\centerline{\includegraphics[width=60mm]{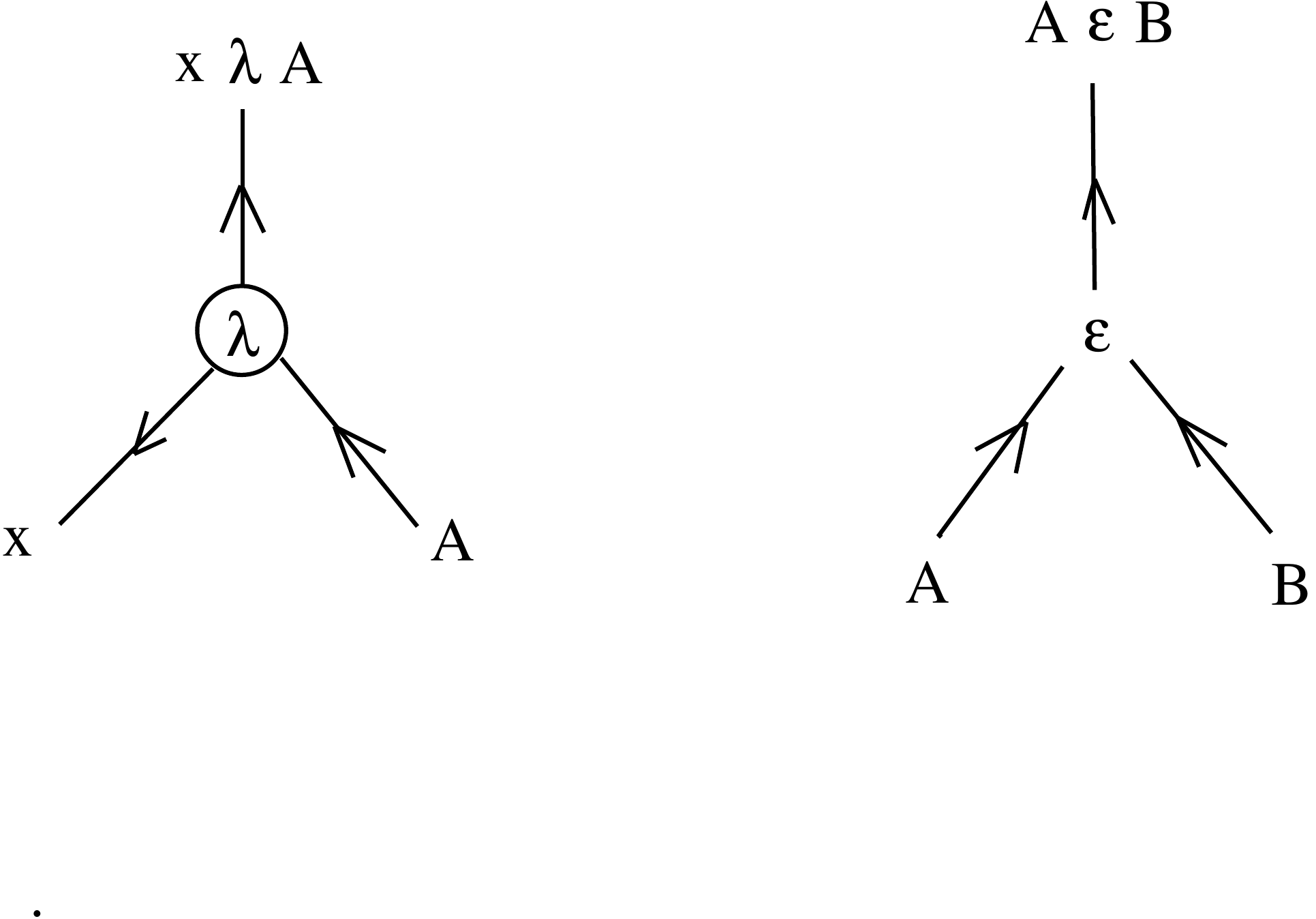}}

The application operation, which will appear in the syntactic tree notation as the gate 
$\curlywedge$,  is defined as 
$$ (A,B) \mapsto AB = A 1 B$$ 
where $A$, $B$ are terms in $\lambda$-Scale calculus and $1$ is the neutral element of the group $\Gamma$.

In this calculus we define another operation, which corresponds to the $\displaystyle \bar{\varepsilon}$ gate, by the formula: 
$$\displaystyle B\bar{\varepsilon} A = (y \lambda A)\varepsilon B$$ 
with $y$ a fresh variable (see definition of "dilation operation", section 2.1, definition 2.2 \cite{lambdascale}). Thus, graphically, the elementary tree $\displaystyle \bar{\varepsilon}$ appears as a composite. 

\vspace{.5cm}

\centerline{\includegraphics[width=120mm]{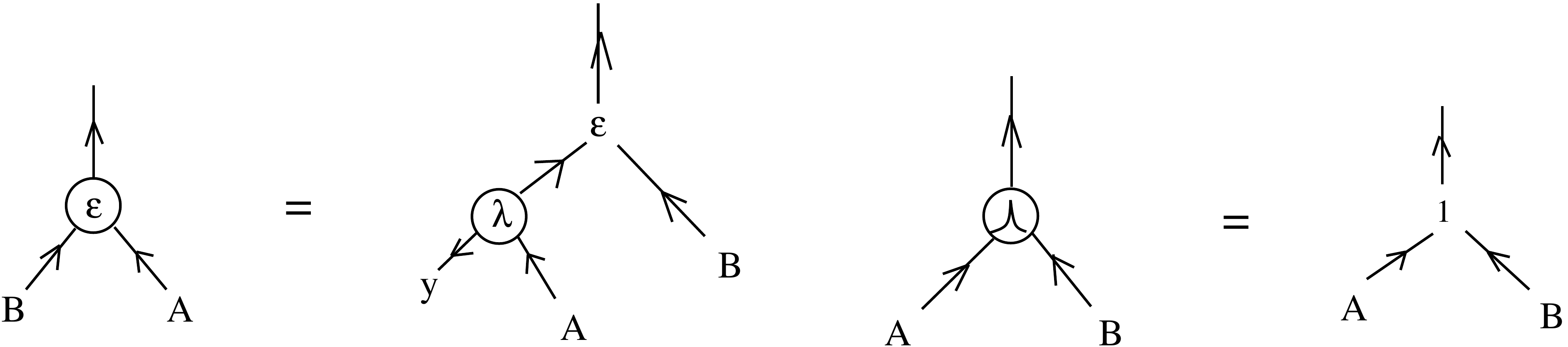}}

\vspace{.5cm}

 In proposition 3.2 \cite{lambdascale} we prove that 
$$B \varepsilon A \, \leftrightarrow \, A \bar{\varepsilon} \left( AB \right)$$
by using a (ext1) move and a ($\beta$*) move.   In theorem 3.5 \cite{lambdascale} we prove that the $\beta$-reduction from lambda calculus extends to terms in $\lambda$-Scale calculus, by using the ($\beta$*) move and (ext1) (which corresponds to an extension of 
$\eta$-reduction to terms in $\lambda$-Scale calculus).

\paragraph{Transformation of syntactic trees in $\lambda$-Scale calculus into graphs in $GRAPH$.} Up to the use of (ext1) move and ($\beta$*) move, we can transform a syntactic tree of a term in $\lambda$-Scale calculus into a syntactic tree of an (equivalent via the transitive closure of moves) term written only with the help of operations $\lambda$, $\curlywedge$ and $\displaystyle \bar{\varepsilon}$. For these terms we use the set of moves: ($\beta$)-reduction, and (R1), (R2), (ext1), (ext2) expressed with the help of the operations $\lambda$, $\curlywedge$ and $\displaystyle \bar{\varepsilon}$ (see the proof of theorem 3.4 \cite{lambdascale}, where it is explained how (R1), (R2) and (ext2) are transformed). 

From here, the procedure of transformation of the syntactic tree of a term in lambda calculus into a graph in $GRAPH$ extends verbatim to a procedure of transformation of the mentioned syntactic trees of terms in $\lambda$-Scale calculus. That is because the procedure described in section \ref{constru} eliminates variables and acts only on the $\lambda$ gates, therefore the said procedure apply as well to trees which have also   $\bar{\varepsilon}$ gates. 

Here is a result which is similar to theorem \ref{lambdathm}. 

\begin{theorem}
Let $A \mapsto [A]$ be a transformation of a $\lambda$-Scale term $A$ (written only with operations $\lambda$, $\curlywedge$ and $\displaystyle \bar{\varepsilon}$) into a graph $[A]$ by using (the extension of) the procedure described in section \ref{constru}. Then: 
\begin{enumerate}
\item[(a)] for any term $A$ the graph $[A]$ is in $GRAPH$, 
\item[(b)] if $[A]'$ and $[A]"$ are transformations of the term $A$ then we may pass from 
$[A]'$ to $[A]"$ by using a finite number (exponential in the number of leaves of the syntactic tree of $A$) of (CO-ASSOC) moves,
\item[(c1)] if $B$ is obtained from $A$ by $\alpha$-conversion then we may pass from  $[A]$ to $[B]$ by a finite sequence of (CO-ASSOC) moves,
\item[c2)]  if $B$ is obtained from $A$ by (ext1), (ext2), (R1) or (R2) rules in $\lambda$-Scale, then we may pass from  $[A]$ to $[B]$ by a finite sequence of (CO-ASSOC) moves, and moves (ext1), (ext2), (R1) and (R2) acting on $GRAPH$, 
\item[(d)] let $A, B \in T(X)$ be two terms and $x \in X$ be a variable. Consider the terms 
$x \lambda A$ and $A[x:=B]$, where $A[x:=B]$ is the term obtained by substituting in $A$ the free occurrences of $x$ by $B$.  Then, by one graphic ($\beta$) move in $GRAPH$ applied to  $[(x \lambda A) B]$ we pass to a graph which can be further  transformed into one  of  $A[x:=B]$, via (global FAN-OUT) moves,  (CO-ASSOC) moves and pruning moves. 
\end{enumerate}
\label{lambdascalethm}
\end{theorem} 

\paragraph{Proof.} (a), (b) (c1) and (d) have the same proof as the one for (a), (b) (c), (d) theorem \ref{lambdathm}. For (c2) see the proof of theorem 3.4 \cite{lambdascale} in order to transform the moves (R1), (R2), (ext2) from $\Lambda$-Scale calculus into moves 
expressed with the gates  $\lambda$, $\curlywedge$ and $\displaystyle \bar{\varepsilon}$. The move (ext1) is already expressed with these gates.  The procedure of transformation of $A$ into $[A]$ transforms then these moves into the graphic moves  (ext1), (ext2), (R1) or (R2). \quad $\square$

\section{Outside lambda calculus: crossings}
\label{secbraid}

In this section we use the graphic $\beta$ move in ways which are exterior to lambda calculus. 

\paragraph{Planar $\beta$ moves.} Remark that we may put the graphic ($\beta$) move in the following form: 

\vspace{.5cm}

\centerline{\includegraphics[width=100mm]{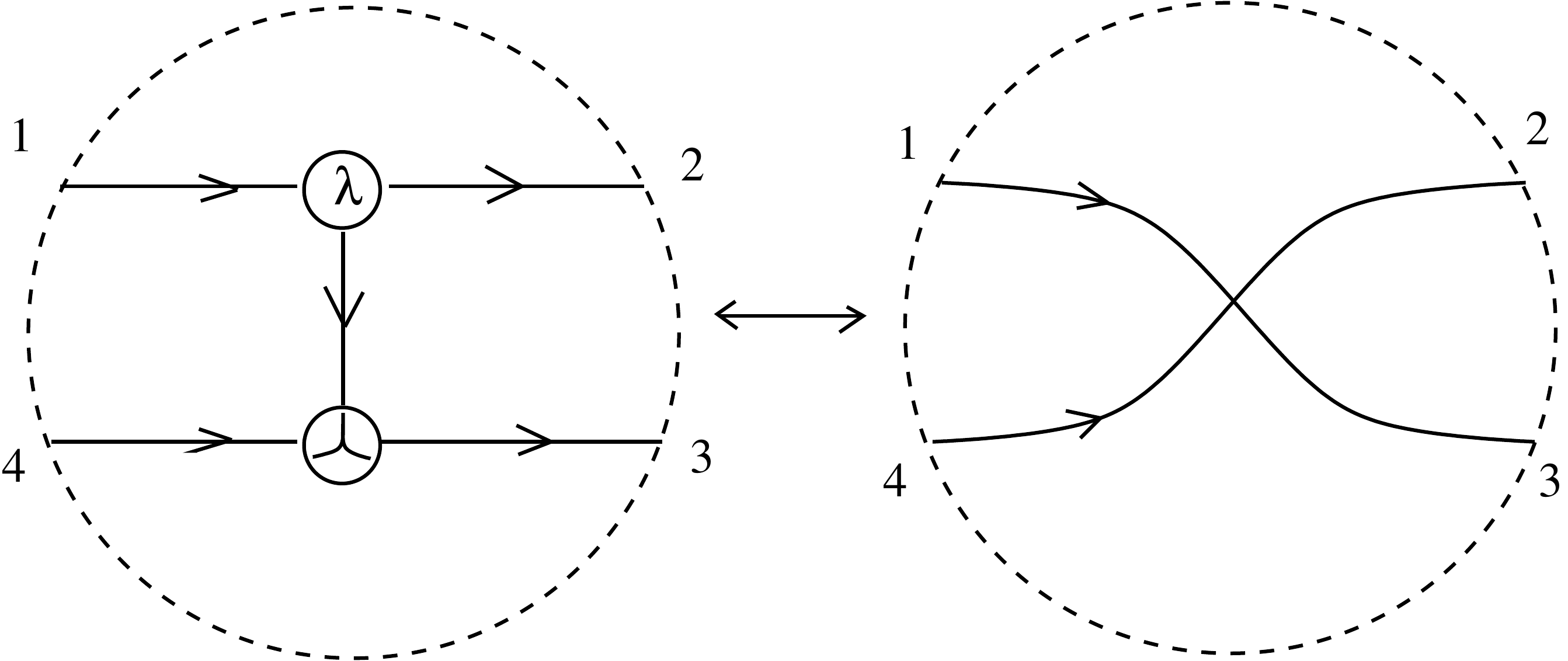}}

\vspace{.5cm}

We use then the graphic ($\beta$) move in order to prove the planar ($\beta$) moves, listed furter. 

\paragraph{($\beta$P1) move.} The particular embedding of the arrow from "1" to "2" is graph-theoretic  not relevant, but we can exploit it in order to prove this move.

\vspace{.5cm}

\centerline{\includegraphics[width=100mm]{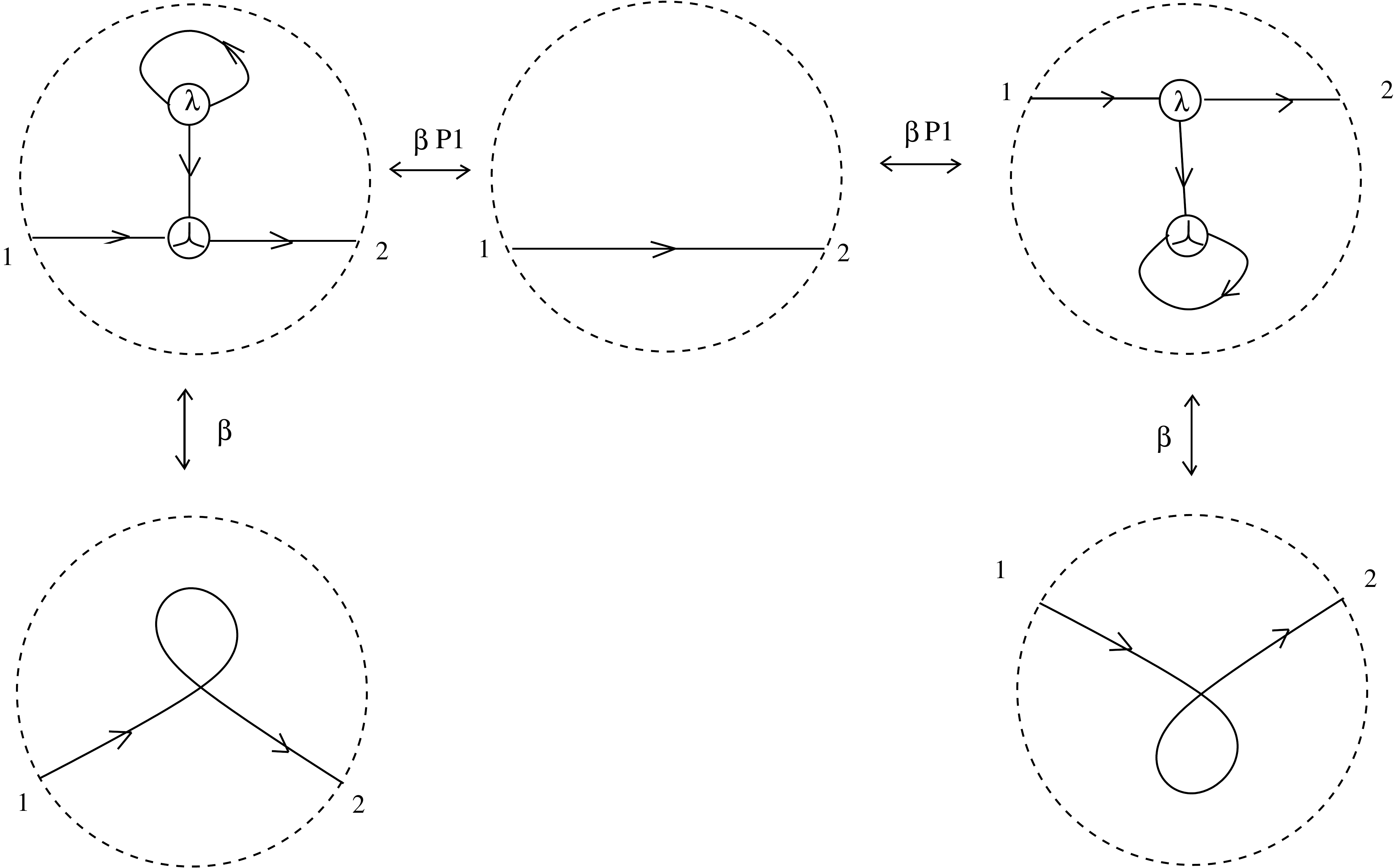}}

\vspace{.5cm}

\paragraph{($\beta$P2) move.} Likewise, it does not matter if the embeddings of arrows from "1" to "2" and from "3" to "4" are crossing or not, but this may be used for proving this move, via two graphic ($\beta$) moves. 

\vspace{.5cm}

\centerline{\includegraphics[width=80mm]{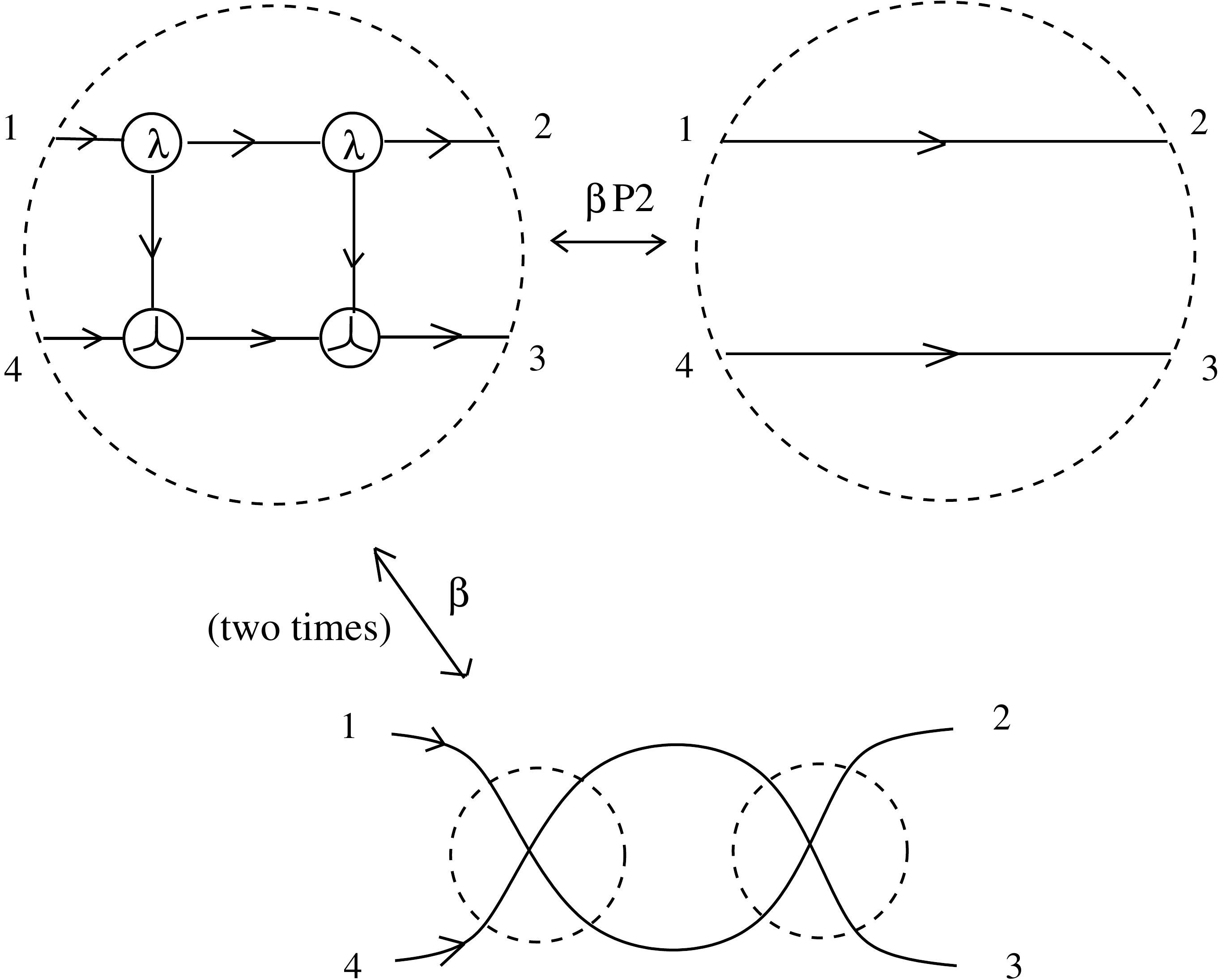}}

\vspace{.5cm}

\paragraph{($\beta$P3) move.} Finally, it does not matter how "1" connects to "4", "6" to "3" and "5" to "2", but we can use three graphic ($\beta$) moves for proving this. 

\vspace{.5cm}

\centerline{\includegraphics[width=130mm]{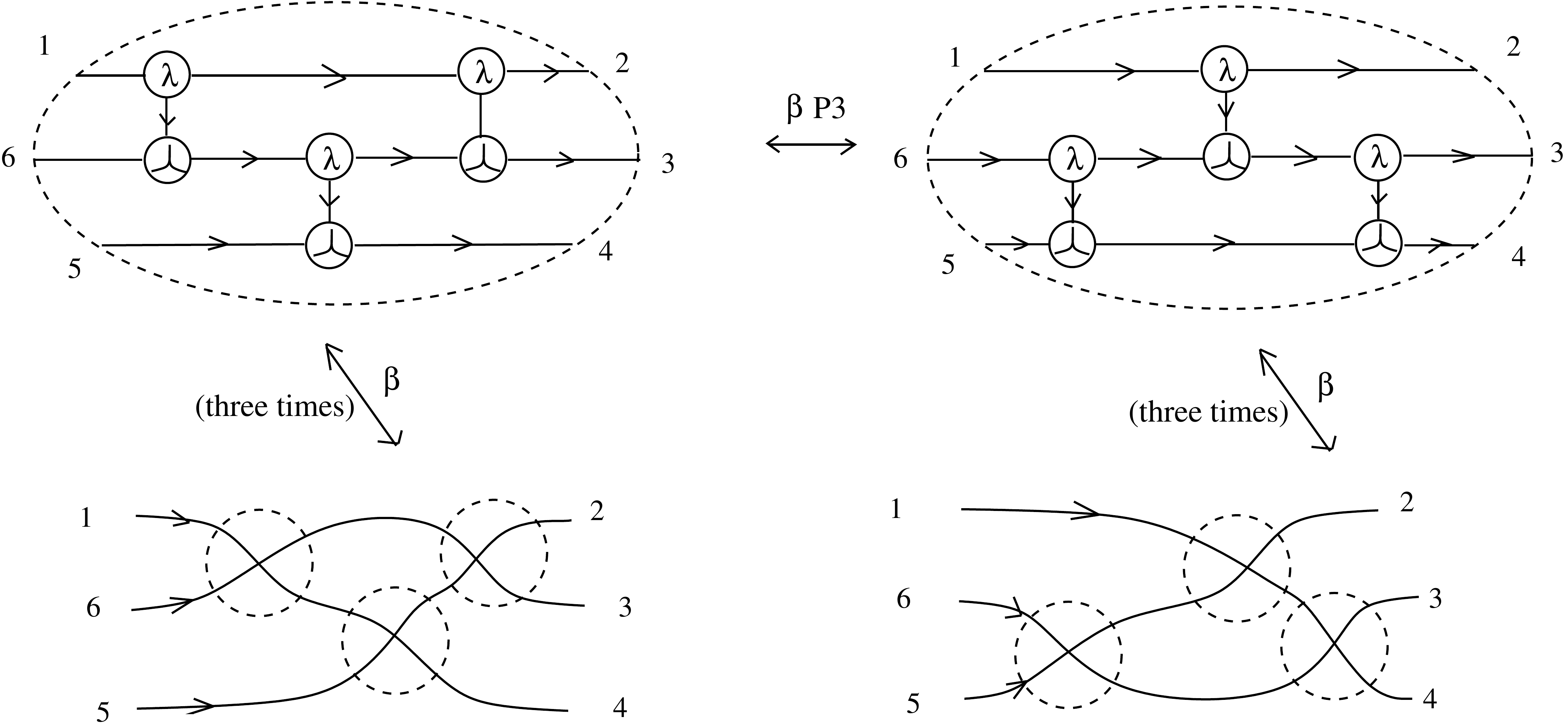}}

\vspace{.5cm}

\paragraph{Braids crossings.} 
In the following figure $\varepsilon \in \Gamma$ is an arbitrary element of the abelian group 
$\Gamma$.

We may "code" a crossing in a braid diagram in the following way. 

\vspace{.5cm}

\centerline{\includegraphics[width=100mm]{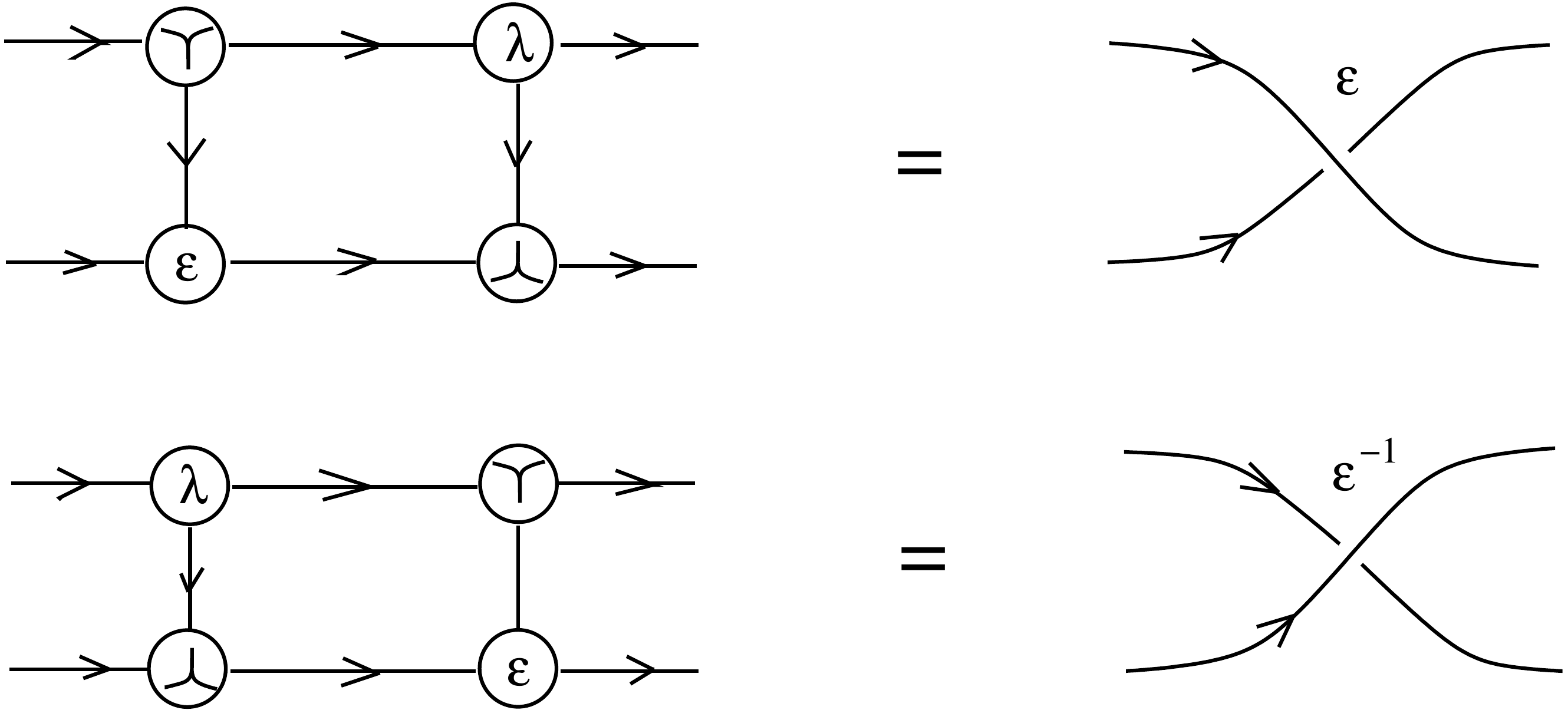}}

\vspace{.5cm}

Let us prove that these crossings are one the inverse of another (this corresponds to the Reidemeister II move for braids).
 
\vspace{.5cm}

\centerline{\includegraphics[width=100mm]{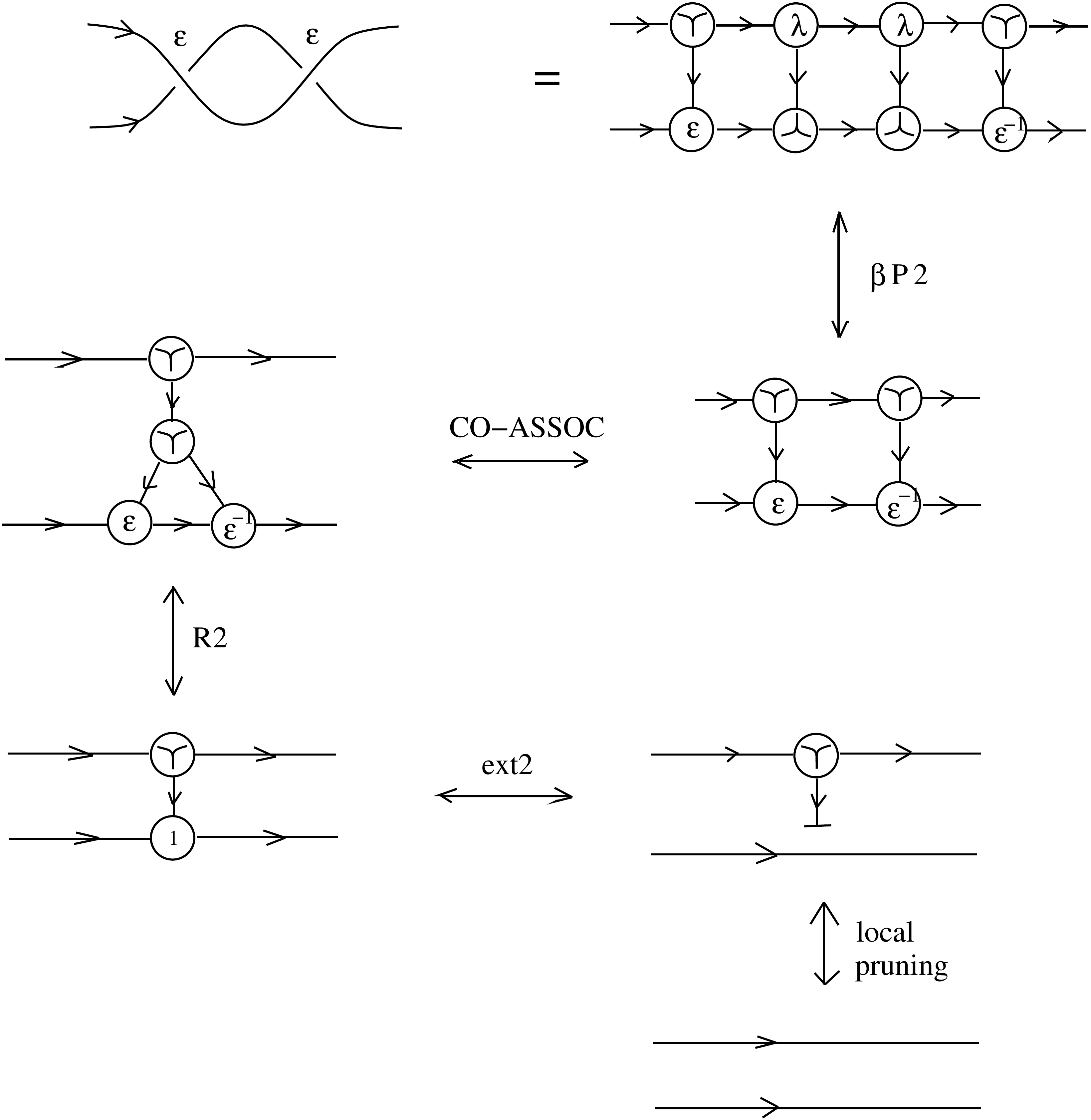}}

\vspace{.5cm}

We keep the investigation of these crossings for another paper, here we remark that for these crossings to satisfy the Reidemeister III move, one needs to add a new move, corresponding to 
what we called "linearity" in the realm of dilation structures, or left distributivity in he frame of emergent algebras.


\begin{thebibliography}{99}


\bibitem{buligadil1} M. Buliga, Dilatation structures I. Fundamentals, {\it 
J. Gen. Lie Theory Appl.},  {\bf 1} (2007),  2, 65-95. 



\bibitem{buligairq} M. Buliga, Emergent algebras, \url{http://arxiv.org/abs/0907.1520}

\bibitem{buligabraided} M. Buliga, Braided spaces with dilations and sub-riemannian symmetric spaces, in: Geometry. Exploratory Workshop on Differential Geometry and its Applications, eds. D. Andrica, S. Moroianu, Cluj-Napoca 2011, 21-35,  
\url{http://arxiv.org/abs/1005.5031}

\bibitem{lambdascale} M. Buliga, $\lambda$-Scale, a lambda calculus for spaces with dilations, 
\url{http://arxiv.org/abs/1205.0139}

\bibitem{buligachora} M. Buliga, Computing with space: a tangle formalism for chora and difference , \url{http://arxiv.org/abs/1103.6007}






\bibitem{fennrourke} R. Fenn, C. Rourke, Racks and Links in codimension two, 
 {\it J. Knot Theory Ramifications},  {\bf 1}  (1992),  no. 4, 343--406



\bibitem{joyce} D. Joyce, A classifying invariant of knots; the knot quandle, 
{\it J. Pure Appl. Alg.}, {\bf 23} (1982), 37-65 

\bibitem{kauf} L. Kauffman, Knot logic. Knots and applications, 1–110, Ser. Knots Everything, 6, World Sci. Publ., River Edge, NJ, 1995

\bibitem{koen} J.J. Koenderink, The brain a geometry engine, {\it Psychological Research}, 
{\bf 52} (1990), no. 2-3, 122-127





\end{thebibliography}
\end{document}